\documentclass[]{aastex631}

\usepackage{comment}
\usepackage{color}
\shorttitle{Polaris}
\shortauthors{Shimoikura et al.}
\graphicspath{{./}{figures/}}

\begin{document}

\title{Velocity Structure and Molecular Formation in Polaris Molecular Cloud}

\correspondingauthor{Tomomi Shimoikura}
\email{ikura@otsuma.ac.jp}

\author[0000-0002-1054-3004]{Tomomi Shimoikura}
\affiliation{Otsuma Women's University Chiyoda-ku, Tokyo 102-8357, Japan}
\affiliation{National Astronomical Observatory of Japan, Mitaka, Tokyo 181-8588, Japan}

\author[0000-0001-8058-8577]{Kazuhito Dobashi}
\affiliation{Tokyo Gakugei University Koganei, Tokyo 184-8501, Japan}

\author[0000-0001-5431-2294]{Fumitaka Nakamura}
\affiliation{National Astronomical Observatory of Japan, Mitaka, Tokyo 181-8588, Japan}
\affiliation{Department of Astronomical Science, School of Physical Science, SOKENDAI (The Graduate University for Advanced Studies), Osawa, Mitaka, Tokyo 181-8588, Japan}

\author[0000-0003-4402-6475]{Kotomi Taniguchi}
\affiliation{National Astronomical Observatory of Japan, Mitaka, Tokyo 181-8588, Japan}




\begin{abstract}
We present a wide-field $(60\arcmin \times 30\arcmin)$ study of a dense region within the Polaris Flare, 
hereafter referred to as the `Polaris molecular cloud', using $^{12}$CO, $^{13}$CO, and C$^{18}$O ($J=1-0$) observations at $20\arcsec$ resolution, obtained with the Nobeyama 45 m Radio Telescope. 
The analysis reveals molecular gas formation occurring at column densities up to $\sim10^{21}$ cm$^{-2}$, evidenced by an anti-correlation between $\textsc{Hi}$ and CO distributions, indicating active atomic-to-molecular gas conversion.
We found a threshold column density for molecular formation at $\sim5\times10^{20}$ cm$^{-2}$, which is common among more evolved molecular clouds.
The CO-to-H$_2$ conversion factor, $X_{\rm CO}$, was found to be $0.7 \times 10^{20}$ H$_2$ cm$^{-2}$ (K km s$^{-1})^{-1}$, lower than the solar neighborhood average.
Our chemical models estimate the cloud's age to be $\sim10^{5}-10^{6}$ years, suggesting an early stage of molecular cloud evolution. 
This interpretation is consistent with the observed low $X_{\rm CO}$ factor.
While virial analysis suggests that the entire cloud is gravitationally unbound, we identified several filamentary structures extending from the main cloud body.
These filaments show systematic velocity gradients of $0.5-1.5$ km s$^{-1}$ pc$^{-1}$, and analysis of the velocities shows that the molecular gas within them is falling toward the main cloud body, following a free-fall model. 
This suggests ongoing mass accumulation processes through the filaments, demonstrating that gravitational processes can be important even at column densities of $\sim10^{21}$ cm$^{-2}$.
\end{abstract}




\section{Introduction} \label{sec:intro}

The evolution of molecular clouds from low-density, from atomic gas to dense, star-forming regions remains a fundamental question in astrophysics. 
Understanding this transition is crucial to reveal the mechanisms of star formation and the life cycle of interstellar matter. High-latitude cirrus clouds, with their low densities and negligible star formation, offer unique laboratories for studying the early stages of this evolution. 
The Polaris Flare, a region of high-latitude cirrus clouds located at an estimated distance of 350 pc \citep{Panopoulou}, has been a subject of particular interest in this field \citep[e.g.,][]{Heithausen1993}.

The Polaris Flare is characterized by low visual extinction \citep[$A_{\rm V} < 1$ mag,][]{Zagury, Bernard1999, Dobashi2005, Dobashi2011}. 
It has been thought to be dominated by turbulent motion and mostly gravitationally unbound, which is believed to be the cause of the low star formation efficiency in this region \citep[e.g.,][]{Magnani1985, Hearty1999}. 
These characteristics suggest that the Polaris Flare could be in an early stage of molecular cloud evolution, where gravitational effects have just begun to influence cloud dynamics.

\begin{figure}
\begin{center}
\includegraphics[scale=.2]{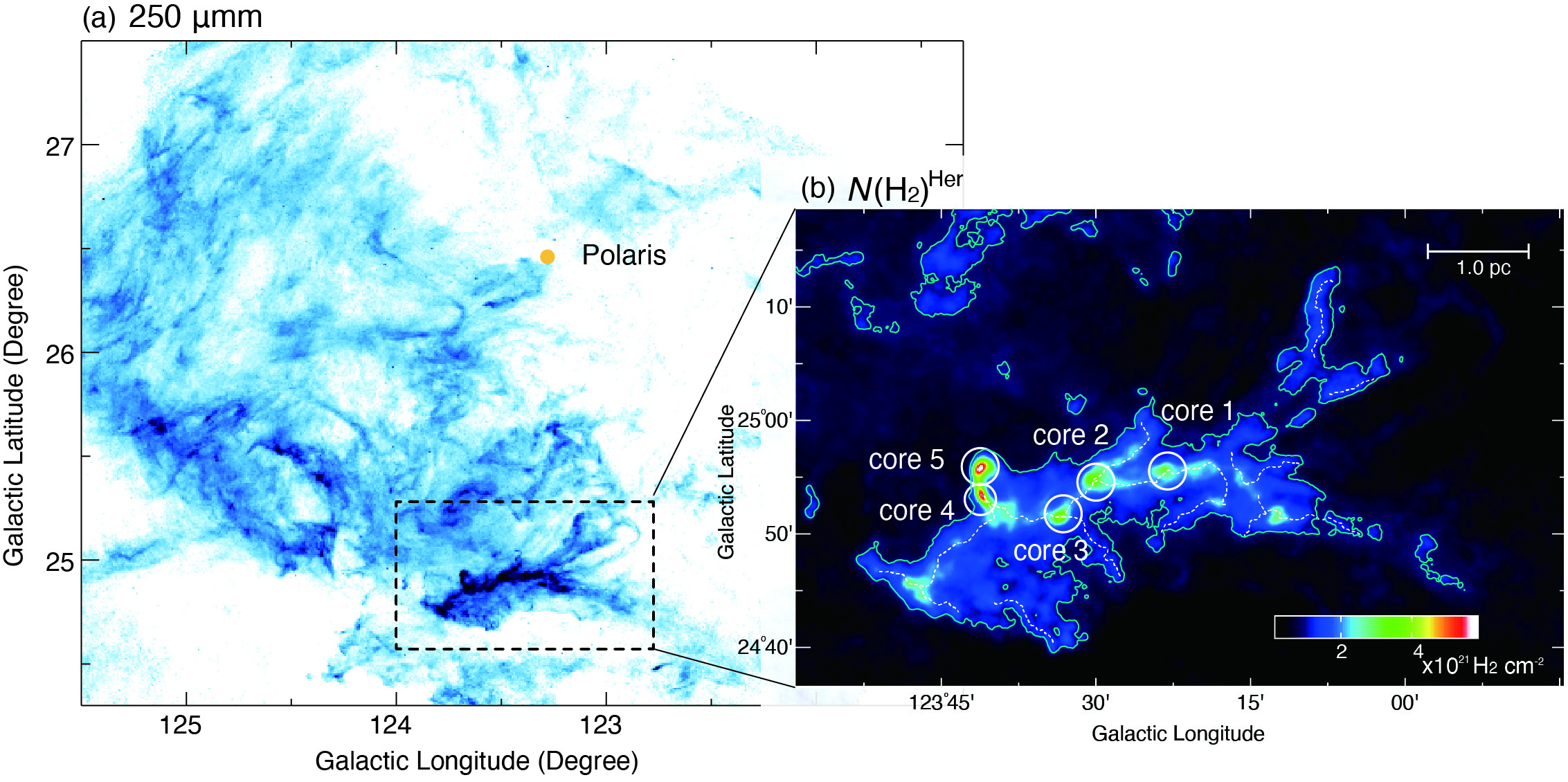}
\caption{
(a) The Polaris Flare from $Herschel$/SPIRE $250\micron$m observations \citep{Menshchikov2010,Miville2010,Ward2010,Andre2010}.
The orange dot indicates the position of Polaris, and
the dotted box highlights the region of the Polaris molecular cloud observed in this study. 
(b) The column density map of the observed region derived from the $Herschel$ observations.
The color scale represents H$_2$ column density in units of $10^{21}$ cm$^{-2}$.
The blue contours indicate the column density of $1\times10^{21}$ cm$^{-2}$.
White circles denote the five cores (numbered 1-5) identified by \cite{Ward2010}, and
the white dotted line delineates the filament structures observed in the $Herschel$ data \cite[e.g.,][]{Andre2013}.
\label{fig:all}}
\end{center}
\end{figure}

Far-infrared observations by the $Herschel$ Space Telescope have revealed a complex structure (Figure \ref{fig:all}a), including more than 30 filamentary structures with a width of about 0.1 pc \cite[e.g.,][]{Menshchikov2010, Miville2010, Ward2010, Andre2010, Andre2013}. 
These filaments, now recognized as fundamental to star formation and mass accumulation in molecular clouds \cite[e.g.,][]{Arzoumanian2019},
contain over 300 embedded prestellar cores \cite[e.g.,][]{Andre2010}.
Although the gravitational equilibrium of these cores remains uncertain, the presence of these prestellar cores indicates that some parts of the Polaris Flare have evolved to densities characteristic of the earliest stages of star formation processes.

This study focuses on a relatively dense region of $\sim 60\arcmin \times 30\arcmin$ (corresponding to approximately 6 pc  $\times$ 3 pc at the distance of 350 pc) within the Polaris Flare (Figure \ref{fig:all} b), which we refer to as the ``Polaris molecular cloud". 
Within this cloud, five high-density cores have been identified, including two cores (cores 4 and 5, also known as the MCLD123.5+24.9 cores) that have been studied in detail \citep{Grossmann1992, Heithausen1999, Shimoikura2012}. 
These cores have densities of $n$(H$_2) > 10^{5}$ cm$^{-3}$, corresponding to column densities of $N$(H$_2) > 10^{21}$ cm$^{-2}$ as seen in the color scale of Figure \ref{fig:all} (b).
They are rich in molecules such as CCS and HC$_3$N, suggesting they are in an early stage of chemical evolution prior to star formation \citep{Suzuki1992, Shimoikura2012}.

The CO-to-H$_2$ conversion factor, known as the $X_{\rm CO}$ factor, in the solar neighborhood is typically 
$2.0\times10^{20}$ H$_2$ cm$^{-2}$ (K km s$^{-1})^{-1}$ \citep[e.g.,][]{Strong1996, Dame2001}. 
However, previous studies have reported a significantly lower value of $0.4\times10^{20}$ H$_2$ cm$^{-2}$ (K km s$^{-1})^{-1}$ for the broader Polaris Flare region \citep{Heithausen1990}. 
The $X_{\rm CO}$ factor for the Polaris molecular cloud itself, a small part of the broad Polaris Flare, has not yet been clearly measured, limiting our understanding of the cloud's specific properties and evolutionary state.

The filamentary structures evident in Figure \ref{fig:all} (b) are characteristic features of molecular clouds and are thought to play a crucial role in the accumulation of material that may lead to star formation.
In molecular clouds with a higher column density ($10^{22} - 10^{23}$ cm$^{-2}$),
infall motions along filaments have been observed \citep[e.g.,][]{Kirk2013,her1,Chen2019, Shimoikura2016, Shimoikura2022}.
Recent numerical simulations support these observational findings and have advanced our understanding of molecular cloud evolution. 
For instance, \cite{Smith} emphasize the role of gravity in the early stages of molecular cloud formation, while \cite{Enrique2019} show that molecular clouds undergo global gravitational collapse, during which high-density structures necessary for star formation are created.
However, such phenomena have not been thoroughly investigated in low-density environments like the Polaris molecular cloud. 
This gap in our knowledge is particularly significant for developing a complete picture of molecular cloud evolution. 
Furthermore, the Polaris molecular cloud's location in a high-latitude region, away from the galactic plane, minimizes the effects of contamination, making it an ideal target for studying intrinsic cloud dynamics.

Given these characteristics, the Polaris molecular cloud provides a unique laboratory for studying the early stages of molecular cloud evolution.
Using the Nobeyama 45 m telescope, we mapped the $J=1-0$ transition lines of $^{12}$CO, $^{13}$CO, and C$^{18}$O across the $\sim 60\arcmin \times 30\arcmin$ region containing five dense cores (Figure \ref{fig:all}(b)). 
Combined with \textsc{Hi} 21 cm line data from the HI4PI survey \citep{HI4PI}, 
we investigate the molecular gas distribution, velocity structure, and $X_{\rm CO}$ factor to understand the cloud's formation process and evolutionary stage.

This paper is structured as follows:
Section \ref{sec:obs} explains the observations, Section 3 presents the main results, 
Section 4 provides discussion of the results, focusing on the evolutionary stage and dynamical state of the Polaris molecular cloud. 
Finally, Section \ref{sec:conclusions} summarizes our findings.

\section{Observations and Data} \label{sec:obs}

\subsection{NRO 45m Observations}

Observations with the NRO 45 m telescope were conducted over two periods, January to April in both 2022 and 2023, totaling 70 hours.
We focused on a $\sim 60\arcmin \times 30\arcmin$ region in the dust distribution shown in Figure \ref{fig:all} (b).
The angular resolution of the telescope (HPBW) is $\sim15\arcsec$ at 100 GHz. 
The observations utilized the FOREST receiver \citep{Minamidani} and the SAM45 spectrometer.
The spectral window mode was employed to simultaneously acquire the $^{12}$CO, $^{13}$CO, and C$^{18}$O data. 
For each molecular emission line, we used the rest frequency taken from \cite{Lovas}.
The velocity resolution of the spectrometer was set to 0.05 km s$^{-1}$.
We conducted the observations using the On-The-Fly (OTF) technique following the procedure outlined by \cite{Sawada}.
Pointing was achieved with the H40 receiver towards the SiO Maser V-Cam at 1.5-hour intervals, maintaining a pointing accuracy within $3\arcsec$ throughout the observations. 

Data calibration was performed using NOSTAR \citep{Sawada}, a software package developed by NRO. 
The conversion factor from antenna temperature ($T_a^{*}$) to brightness temperature ($T_{\rm{mb}}$) was determined using the annual main beam efficiencies measured by NRO. 
Subsequently, the data were resampled to a grid with a spacing of $7.5\arcsec$ 
and smoothed to a velocity resolution of 0.1 km s$^{-1}$ with a noise level of 1.0, 0.4, 0.3 K for $^{12}$CO, $^{13}$CO, and C$^{18}$O, respectively.

\subsection{Archival Data}

In this study, we utilized three archival datasets for comparative analysis with the 45m data. 
The first dataset comprises the column density data obtained from the Herschel Gould Belt Survey (HGBS) observations. 
This dataset presents the total hydrogen nuclei column density, 
expressed in units equivalent to molecular hydrogen, $N(\rm{H}_{2})$, 
derived from the analysis of  $160 \micron$ far-infrared data obtained by the Herschel Space Telescope in the Polaris Flare region, as detailed on the HGBS website\footnote {http://www.herschel.fr/cea/gouldbelt/en/}. 
We refer to this dataset as $N(\rm{H}_{2})^{Her}$. 
Note that $N(\rm{H}_{2})^{Her}$ includes all the hydrogen nuclei both in the molecular and atomic forms.
The angular resolution of this data is $18\arcsec$.

The second dataset is the \textsc{Hi} 21 cm emission line data obtained by the HI4PI survey \citep{HI4PI}. 
This dataset allows us to explore the distribution of neutral hydrogen atoms in the Polaris molecular cloud. The angular resolution of this data is $16\arcmin$ due to the low rest frequency (1.4 GHz). The velocity resolution is approximately 1.3 km s$^{-1}$.

The third dataset is the $^{12}$CO($J=1-0$) data obtained by the 1.2 m radio telescopes \citep{Dame2001}. The angular resolution is 8.4$\arcmin$.

\section{Results} \label{sec:results}

\subsection{Spatial Distributions of the molecular cloud} \label{sec:ii}

\begin{figure}
\begin{center}
\includegraphics[scale=.2]{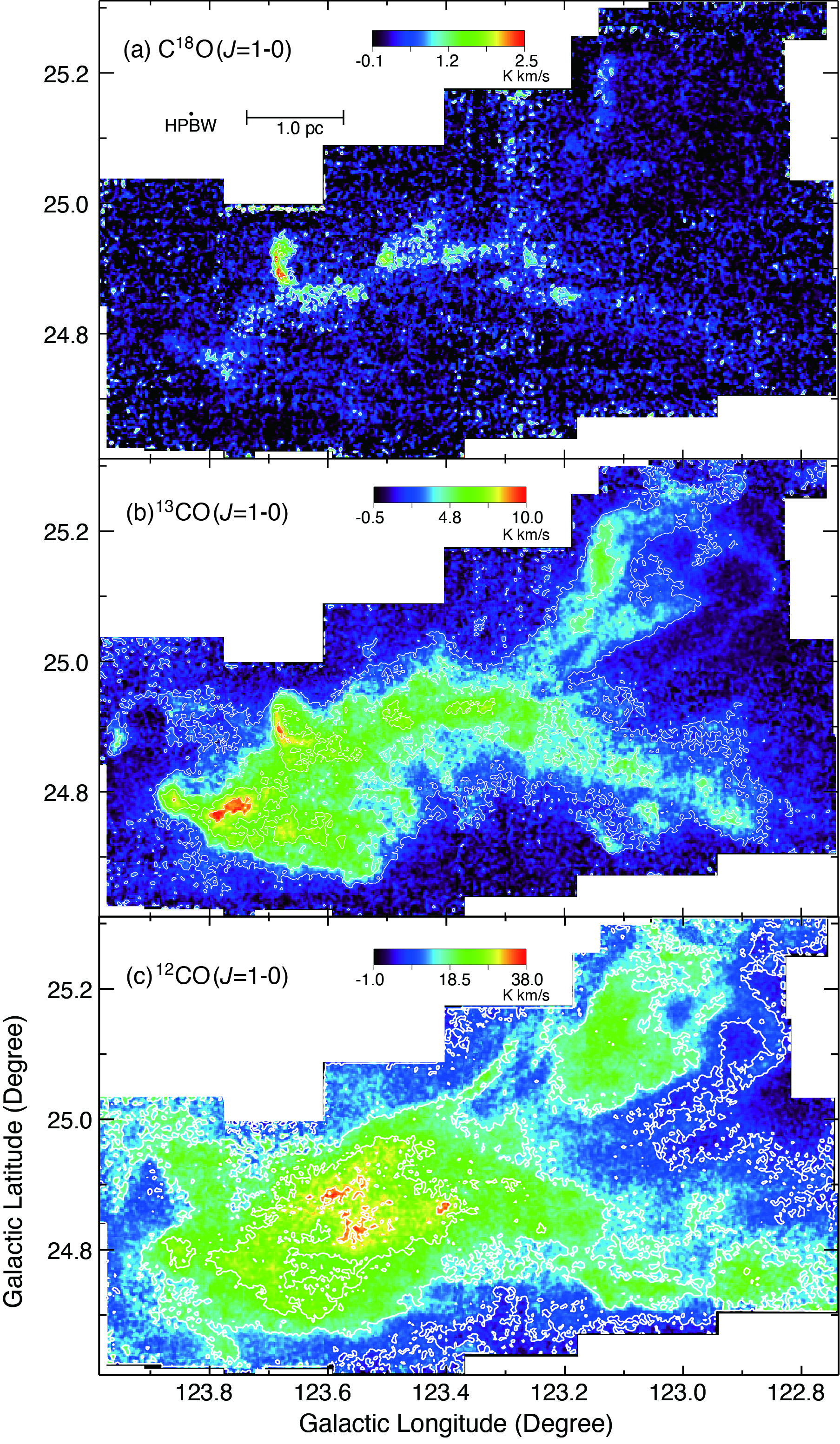}
\caption{
Integrated intensity maps of the molecular emission lines of (a)C$^{18}$O, (b)$^{13}$CO, and (c)$^{12}$CO.
Contour levels start and increment by 0.8, 2.0, and 8.0 K km s$^{-1}$ for C$^{18}$O, $^{13}$CO, and $^{12}$CO, respectively.
The angular resolution (HPBW) is shown in the bottom right corner of panel (a), and the scale bar indicates 1.0 pc at the assumed distance of 350 pc.
\label{fig:ii}}
\end{center}
\end{figure}

We present the integrated intensity maps of C$^{18}$O, $^{13}$CO, and $^{12}$CO obtained by our observations in Figure \ref{fig:ii}.
The C$^{18}$O emission tends to be weak, ranging from 1 to 2.5 K km s$^{-1}$, 
compared to other molecular clouds, such as those in Taurus.
However, prominent C$^{18}$O detection can be seen around the five cores identified by \citet{Ward2010} in the Polaris molecular cloud.
In contrast, the distribution of $^{13}$CO shows a good correlation
with the $N$(H$_2$)$^{\rm Her}$ distribution derived from the $Herschel$ observations.

\begin{figure}
\begin{center}
\includegraphics[scale=.2]{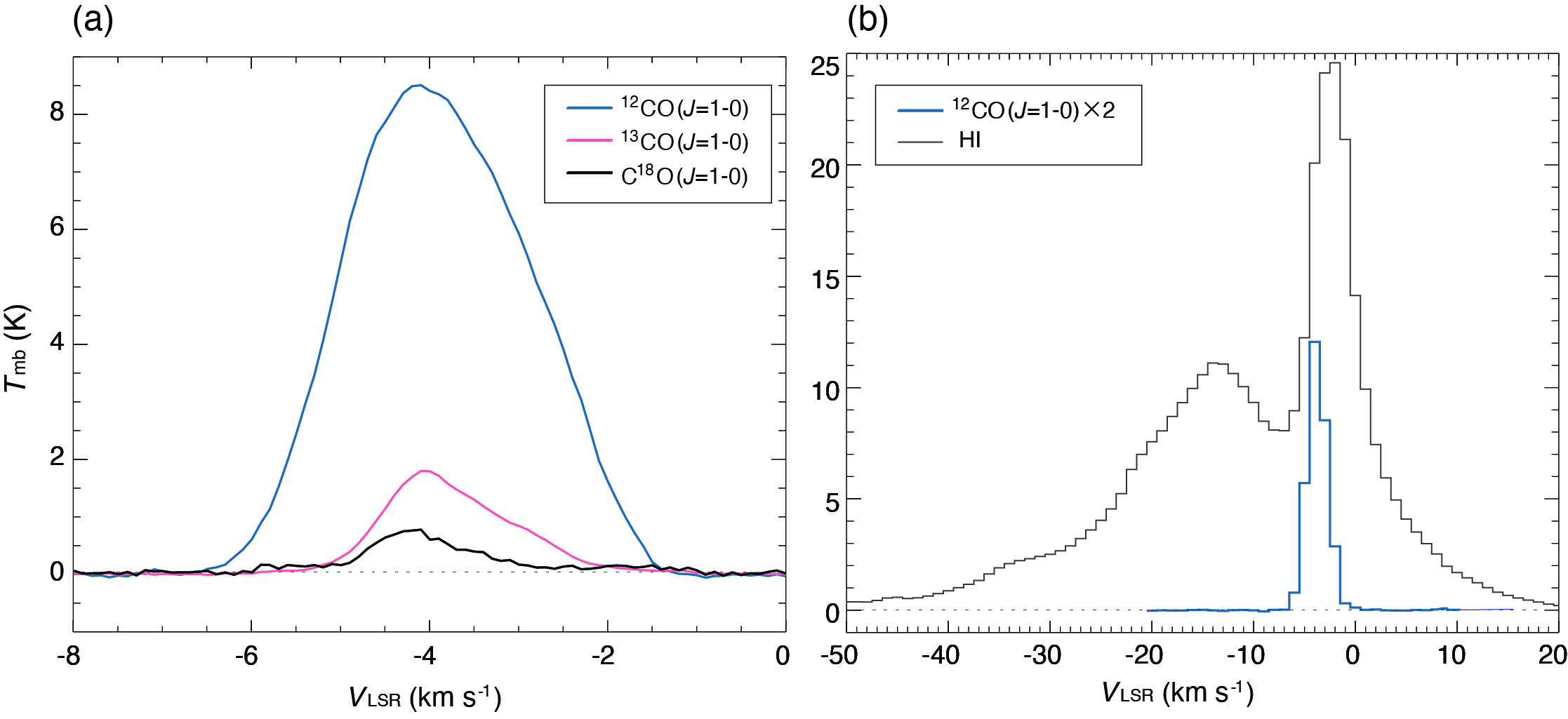}
\caption{
Averaged spectra of (a) the $^{12}$CO, $^{13}$CO, and C$^{18}$O emission lines. The $^{13}$CO and C$^{18}$O emission lines are obtained from pixels above $3\sigma$ noise level, and the $^{12}$CO emission line is obtained from pixels above $5\sigma$ noise level. (b) The $^{12}$CO and $\textsc{Hi}$ 21 cm emission lines within the observation area of Figure \ref{fig:ii}. 
The $^{12}$CO spectrum is smoothed to the same velocity resolution as the $\textsc{Hi}$ 21 cm data.
\label{fig:spe}}
\end{center}
\end{figure}

To investigate the spatial distribution of neutral atomic hydrogen in the vicinity of the Polaris molecular cloud, 
we created an integrated intensity map of the $\textsc{Hi}$ 21 cm emission line, $W_{\textsc{Hi}}$. 
Figure \ref{fig:spe} shows the averaged spectra of the C$^{18}$O, $^{13}$CO, $^{12}$CO, and $\textsc{Hi}$ emission lines in the observed region.
Panel (a) presents the averaged spectra measured above $3\,\sigma$ (for C$^{18}$O and $^{13}$CO) and $5\,\sigma$ (for $^{12}$CO) of their respective map noise levels. 
The noise levels for C$^{18}$O, $^{13}$CO, and $^{12}$CO are 0.2 K km s$^{-1}$, 0.4 K km s$^{-1}$, and 0.8 K km s$^{-1}$, respectively.
Panel (b) shows the average spectra of $^{12}$CO and $\textsc{Hi}$ over the entire $^{12}$CO observed area. 
The three CO emission lines appear in the velocity range of $-7.0<V_{\rm LSR} < -1.0$ km s$^{-1}$ within our observed velocity range of $-20.0<V_{\rm LSR}<15.0$ km s$^{-1}$ as shown in panel (a).
In panel (b), the $\textsc{Hi}$ emission line is observed in the velocity range of $-50.0<V_{\rm LSR}< 20.0$ km s$^{-1}$, exhibiting two velocity components, with one component overlapping the velocity range of the CO emission lines.

\begin{figure}
\begin{center}
\includegraphics[scale=.2]{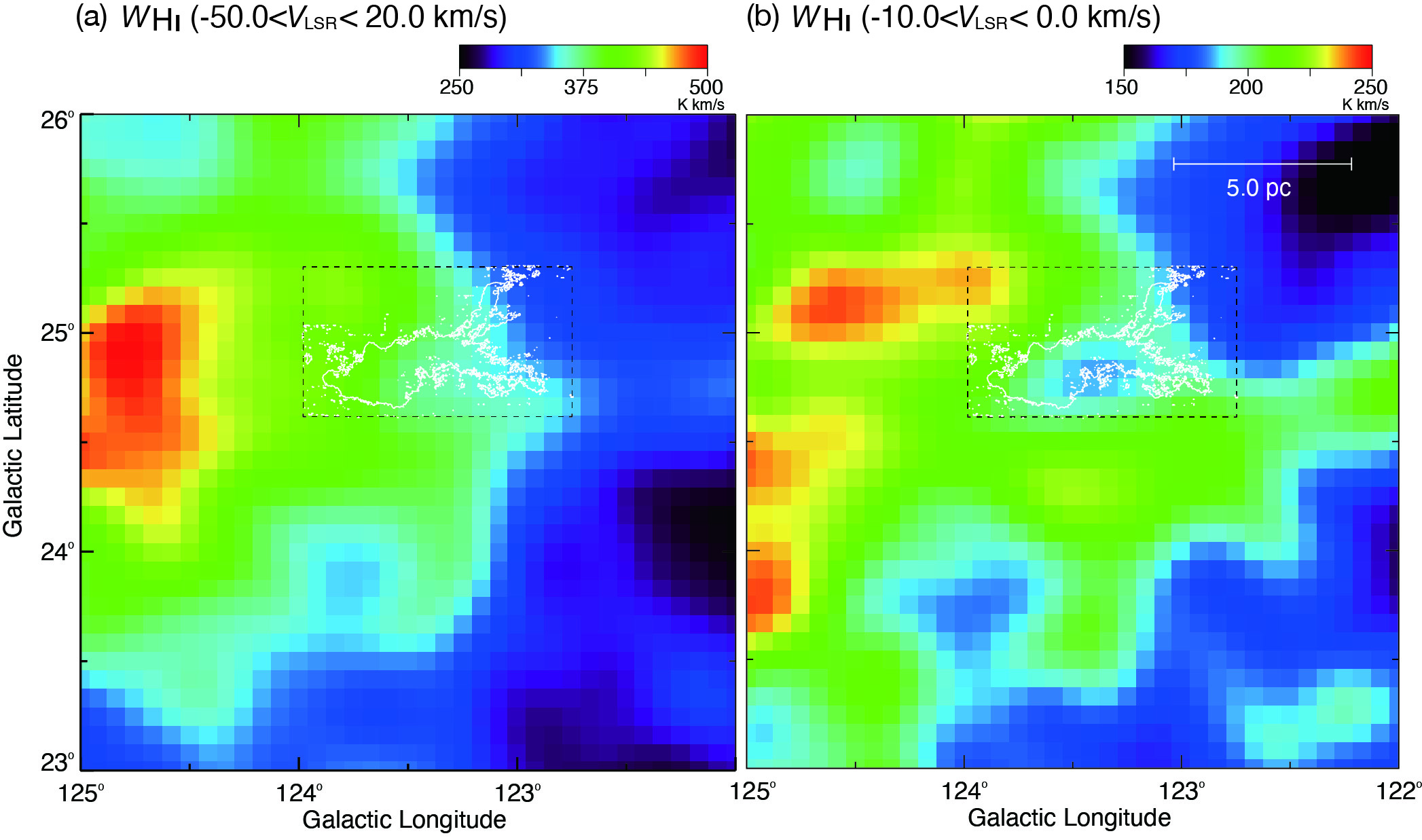}
\caption{
Distribution of $\textsc{Hi}$ 21 cm emission line \citep{HI4PI} around the Polaris molecular cloud. 
The integrated velocity range is indicated in the panel title.
Contours represent the integrated intensity of $^{13}$CO, with a level of 3.0 K km s$^{-1}$. The black dotted line outlines the observed area in this study.
\label{fig:NHI}}
\end{center}
\end{figure}

Figure \ref{fig:NHI} illustrates the distribution of $W_\textsc{Hi}$ (color scale)
shown with the $^{13}$CO map (contours).
In the figure, panel (a) shows $W_\textsc{Hi}$ created by integrating the emission over the velocity range $-50.0<V_{\rm LSR}<20.0$ km s$^{-1}$,
while panel (b) exhibits $W_\textsc{Hi}$ created by integrating the emission over the velocity range $-10.0<V_{\rm LSR}< 0.0$ km s$^{-1}$.
We cannot definitively determine the specific velocity range of the $\textsc{Hi}$ emission associated with the Polaris molecular cloud due to the broad $\textsc{Hi}$ line width and potential contamination from unrelated $\textsc{Hi}$ gas along the line of sight.
The anti-correlation between the intensities of the $\textsc{Hi}$ and $^{13}$CO emission suggests that the $\textsc{Hi}$ emission in this velocity range is associated with the molecular cloud.

\begin{figure}
\begin{center}
\includegraphics[scale=.2]{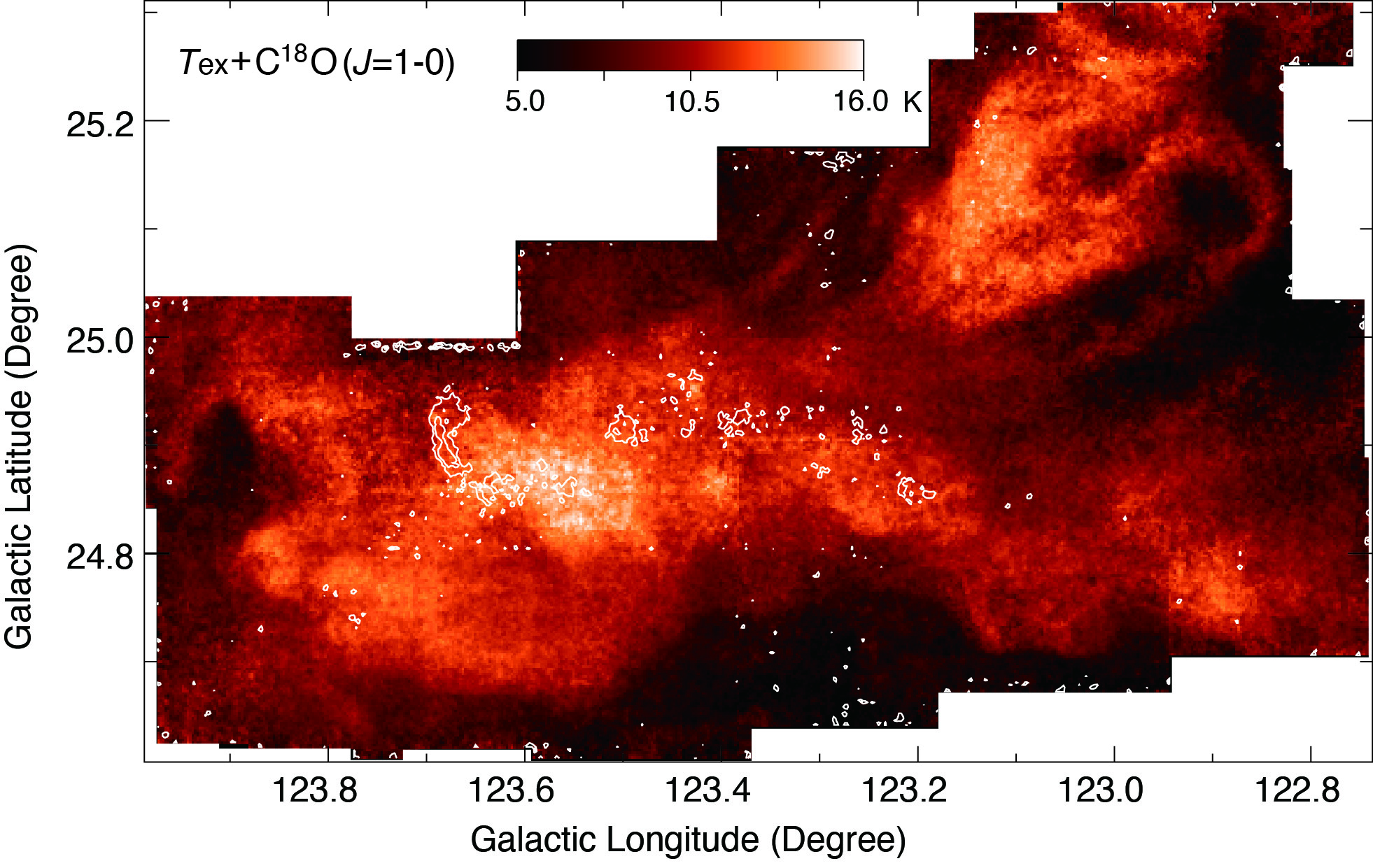}
\caption{
Distribution of excitation temperature $T_ {\rm ex}$ derived from $^{12}$CO, with contours indicating the integrated intensity of C$^{18}$O.
\label{fig:tex}}
\end{center}
\end{figure}

To analyze the anti-correlation quantitatively, we derived the column densities of $\textsc{Hi}$ and 
the CO isotopologues. 
Under the assumption of local thermodynamic equilibrium (LTE),
the excitation temperature $T_ {\rm ex}$ can be obtained from $T^{X}_ {\rm mb}$ of molecule $X$ ($X$ is the observed molecules), optical depth, $\tau^{X}$,
and the brightness temperature of the cosmic background radiation $T_{\rm bg}$(=2.725 K), by solving the following equation,
\begin{equation}
\label{eq:radiative}
T^{X}_ {\rm mb}=[J_{\rm RJ}(T_ {\rm ex})-J_{\rm RJ}(T_{\rm bg})][1-{\rm exp}(-\tau^{X})],
\end{equation}
where $J_{\rm RJ}(T)=T_{0} /(e^{T_{0} /T}-1)$ with $T_{0}=h\nu /k$. 
The constants $k$, $h$, and $\nu$ denote the Boltzmann constant, the Planck constant, and the rest frequency of the observed emission lines, respectively.
To estimate $T_ {\rm ex}$, we used the $^{12}$CO molecular emission line, assuming that the line is optically thick ($\tau^{^{12}\rm{CO}}\gg 1$). 
The distribution of the derived $T_ {\rm ex}$ is shown in Figure \ref{fig:tex},
with the $T_ {\rm ex}$ around each core being $\sim10$ K, equivalent to the dust temperature estimated from HGBS \citep{Ward2010}. 

The values of $\tau$ of the C$^{18}$O and $^{13}$CO molecular emission lines at each pixel were computed using the derived $T_ {\rm ex}$ and Equation (\ref{eq:radiative}).
$\tau^{^{13}\rm{CO}}$ exceeds 1 in the five dense cores, and $\tau^{\rm{C^{18}O}}$ is much smaller ($\lesssim 0.1$).
With the obtained $\tau^X$ and $T_ {\rm ex}$, we calculated the column density of the observed molecules, $N(X)$, utilizing the equation in the following \cite[e.g.,][]{Mangum2015},
\begin{equation}
\label{eq:column}
N(X)=\frac{3h}{8\pi^{3}}\frac{Q}{\mu^{2}S_{ij}}\frac{e^{Eu/k{T}_{\mathrm{ex}}}}{e^{h\nu/kT_{\mathrm{ex}}}-1}\int \tau^{X} dv,
\end{equation}
where $X=^{13}$CO or C$^{18}$O, and $Q$ represents the partition function approximated as $Q=k\,T_{\rm ex}/h\,B_{0}+1/3$ \citep[e.g.,][]{Mangum2015}. $B_{0}$ denotes the rotational constant of the molecule,
$\mu$ is the dipole moment, $E_{u}$ is the energy of the upper level, and $S_{ij}$ is the intrinsic line strength of the transition for $i$ to $j$ state.
Spectral line parameters were obtained from Splatalogue\footnote{https://splatalogue.online/}.
In deriving the column densities, the value of $T_ {\rm ex}$ was set to $T_ {\rm ex}=10$ K for pixels with $T_ {\rm ex} < 10$ K, because the optically thick assumption for the $^{12}$CO emission may not hold for such regions.

We derived neutral atomic hydrogen column density $N(\textsc{Hi})$ by 
\begin{equation}
\label{eq:NHI}
N(\textsc{Hi})=1.823\times10^{18} W_\textsc{Hi}. 
\end{equation}

\begin{figure}
\begin{center}
\includegraphics[scale=.2]{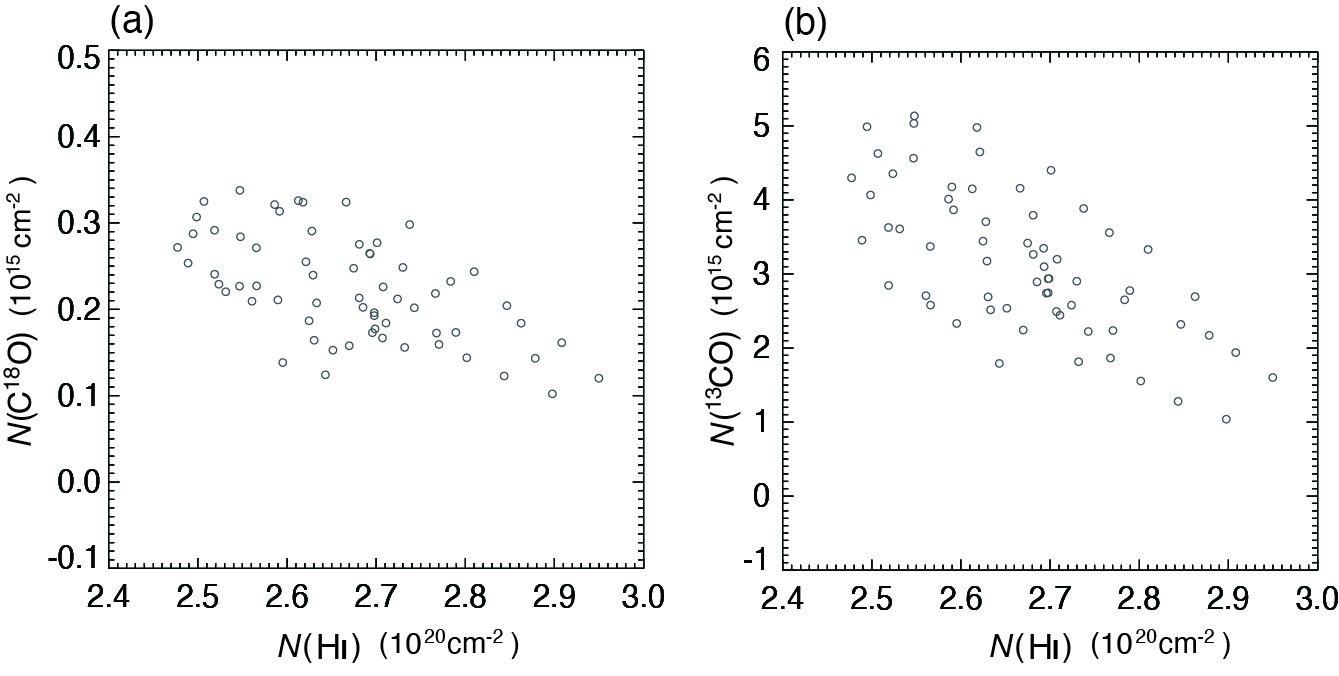}
\caption{
Relationships between (a) $N$(C$^{18}$O) and $N(\textsc{Hi})$, and (b) $N$($^{13}$CO) and $N(\textsc{Hi})$.
All data were smoothed to match the angular resolution of the $\textsc{Hi}$ data.
\label{fig:CO-HI}}
\end{center}
\end{figure}

We subsequently examined the relationships among $N$(C$^{18}$O), $N$($^{13}$CO), and $N(\textsc{Hi})$.
For this, the 45 m data were smoothed to the same angular resolution as the \textsc{Hi} data
to compare directly. Resulting plots are shown in Figure \ref{fig:CO-HI}.
$N(\textsc{Hi})$ exhibits an anti-correlation with both $N$(C$^{18}$O) and $N$($^{13}$CO).
The anti-correlation is particularly evident for $N$($^{13}$CO). 
Since CO is widely recognized as a reliable tracer of molecular hydrogen H$_2$, the observed anti-correlation may represent the conversion of atomic gas to molecular gas 
within the Polaris molecular cloud.


\subsection{Relationship between Column density of the total H-nuclei and Molecular gas}

Following the observed anti-correlation between $\textsc{Hi}$ and CO distributions, 
we investigated the relationship between the column density of the total H-nuclei, $N$(H), and molecular gas in the Polaris molecular cloud.
$N$(H) comprises both atomic hydrogen gas and molecular hydrogen gas, and is expressed as,

\begin{equation}
\label{eq:NH}
N({\rm{H}})=N(\textsc{Hi})+2N({\rm{H}}_2).
\end{equation}

Note that the column density of molecular hydrogen obtained by HGBS, $N(\rm{H}_{2})^{Her}$, is related 
to $N({\rm{H}})$ in the above as $2 N(\rm{H}_{2})^{Her}$=$N$(H).

\begin{figure}
\begin{center}
\includegraphics[scale=.2]{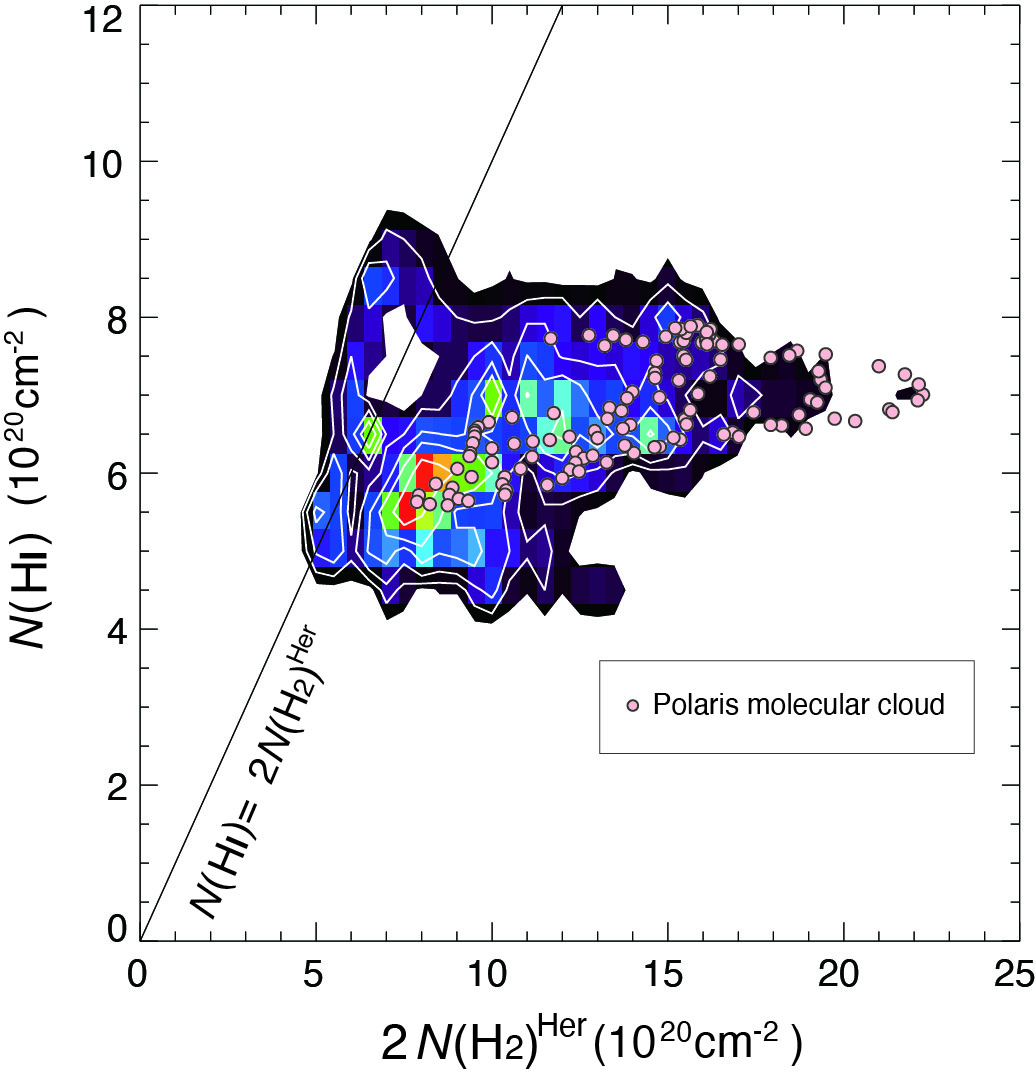}
\caption{
Relationship between $N(\textsc{Hi})$ and $2N(\rm{H}_{2})^{Her}$.
The density plot shows the results for the entire Polaris flare, with the Polaris molecular cloud indicated by pink circles.
The black line reperesents the relationship $N(\textsc{Hi}) = 2N(\rm{H}_{2})^{Her}$.
\label{fig:vsN(H)}}
\end{center}
\end{figure}

Figure \ref{fig:vsN(H)} shows the relationship between $2 N(\rm{H}_{2})^{Her}$ and $N(\textsc{Hi})$. Here, $N(\textsc{Hi})$ is calculated from Equation (\ref{eq:NHI}) 
with $W_\textsc{Hi}$ integrated over the velocity range of $-50.0<V_{\rm LSR}< 20.0$ km s$^{-1}$ (refer to Figure \ref{fig:spe}b).
In the figure, the color-scale density plot represents the entire Polaris Flare region shown in Figure \ref{fig:all}, and the pink circles denote the Polaris molecular cloud. 
While the distribution follows the relationship $N(\textsc{Hi})=2 N(\rm{H}_{2})^{Her}$ in the low column density region, $N(\textsc{Hi})$ becomes flatter and do not increase along with $2 N(\rm{H}_{2})^{Her}$ in
the higher column density region including the Polaris molecular cloud. We suggest that this is due to conversion of atomic hydrogen to molecular hydrogen.

To further investigate the relationship between atomic and molecular hydrogen in the Polaris molecular cloud, we modify Equation (\ref{eq:NH}) to express the column density of molecular hydrogen $N(\rm{H}_{2})$ with $N(\rm{H}_{2})^{Her}$ and $N(\textsc{Hi})$ as
\begin{equation}
\label{eq:NH2}
N({\rm{H}_2})=N({\rm{H}_{2})^{Her}}-\frac{1}{2} N(\textsc{Hi})~.
\end{equation}

\begin{figure}
\begin{center}
\includegraphics[scale=.2]{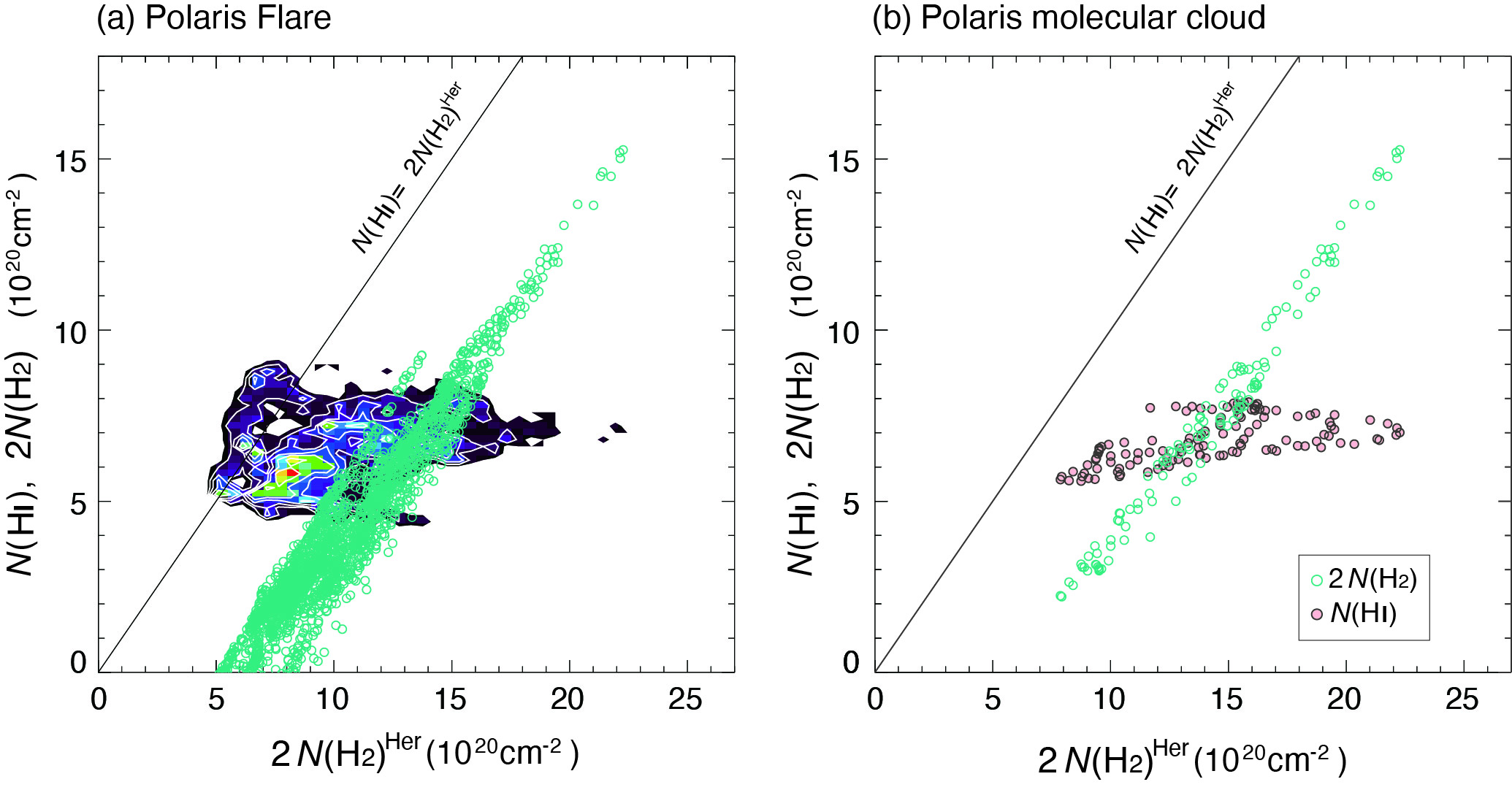}
\caption{
Relation of $N(\textsc{Hi})$ and doubled $N(\rm{H}_{2})$ relative to $2N(\rm{H}_{2})^{Her}$. 
Panel (a) displays the results for the entire Polaris Flare, while panel (b) presents the results for the Polaris molecular cloud only. 
In panel (a), $N(\textsc{Hi})$ is represented by the density plot, 
and doubled $N(\rm{H}_{2})$ is shown by green circles.
In panel (b), $N(\textsc{Hi})$ is represented by pink circles,
and doubled $N(\rm{H}_{2})$ is again shown by green circles.
The black lines in both panels represent the relationship $N(\textsc{Hi}) = 2N(\rm{H}_{2})^{Her}$.
\label{fig:vsN(H)2}}
\end{center}
\end{figure}

In this paper, we regard the gas not tranced by the ${\textsc{Hi}}$ 21cm emission ($N(\textsc{Hi})$)
in the total gas deduced from the dust emission ($N({\rm{H}_{2})^{Her}}$) to be molecular gas ($N(\rm{H_2})$),
and calculated $N({\rm{H}_2})$ using the above equation.
Figure \ref{fig:vsN(H)2}(a) shows the resulting $N({\rm{H}_2})$ shown as a function of $N({\rm{H}_{2})^{Her}}$
for the entire Polaris Flare. The same plots for the Polaris molecular cloud are shown in Figure \ref{fig:vsN(H)2}(b). In these plots, $N(\rm{H}_{2})$ values are doubled to represent their atomic equivalent. 

Figure \ref{fig:vsN(H)2} shows the general trends in the relationships among $N(\textsc{Hi})$, $2N$(H$_2$), and $2 N(\rm{H}_{2})^{Her}$ across a wide range of densities. 
It demonstrates that $2N$(H$_2$) increases along with $2 N(\rm{H}_{2})^{Her}$ (= $N$(H)), while $N(\textsc{Hi})$ remains relatively constant. 
The constant behavior of $N(\textsc{Hi})$ suggests that not all of the interstellar atoms ($\textsc{Hi}$) transform into molecules (H$_2$) in molecular clouds, and a certain fraction of $\textsc{Hi}$ may survive due to cosmic rays or UV radiation. 

Figure \ref{fig:vsN(H)2} also reveals that the molecular hydrogen $2N$(H$_2$) in Polaris starts appearing at the total column density $2 N(\rm{H}_{2})^{Her}\simeq5\times10^{20}$ cm$^{-2}$, which is equivalent to the visual extinction $A_V\sim0.5$ mag, and increases along with the total column density.
This indicates that the total column density $2 N(\rm{H}_{2})^{Her}\simeq5\times10^{20}$ cm$^{-2}$
is needed to initiate molecular formation.
We performed the same analysis for molecular clouds in the Taurus and Orion regions (see the next subsection),
and we actually found similar trends in these regions.
We therefore suggest that there is a common threshold value in the total column density for molecular formation,
and that the threshold value is $2 N(\rm{H}_{2})^{Her}\simeq5\times10^{20}$ cm$^{-2}$.

\subsection{Comparative Analysis of $X_{\rm CO}$ Factor in Molecular Clouds}

To better understand the molecular gas properties of the Polaris molecular cloud, 
we investigated the relationship between $2 N(\rm{H}_{2})^{Her}$ and the integrated $^{12}$CO intensity $W_{\rm CO}$.
To compare with molecular clouds at different evolutionary stages,
we performed a similar analysis for L1495 in the Taurus molecular cloud (referred to as `Taurus'), where low-mass stars have already formed, and the Orion A molecular cloud  (referred to as `Orion A'), where high-mass stars have formed.
$W_{\rm CO}$ data for the Polaris Flare, Taurus, and Orion A were obtained from the $^{12}$CO($J=1–0$) data by \cite{Dame2001}, and $N(\rm{H}_{2})^{Her}$ data for the three regions were taken from the HGBS project \citep[e.g.,][]{her1,her2,her3,her4,her5}.

\begin{figure}
\begin{center}
\includegraphics[scale=.2]{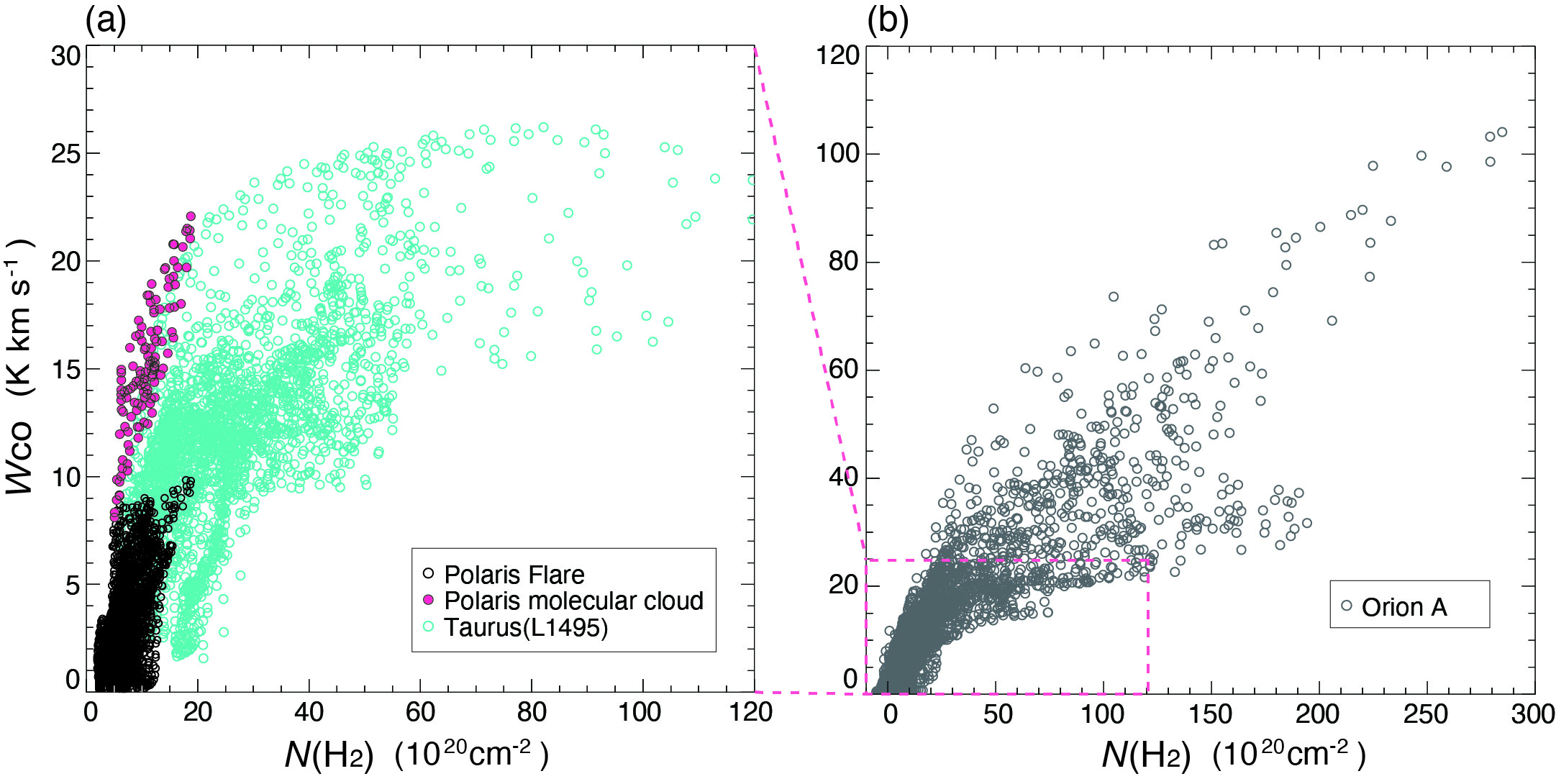}
\caption{
Relationship between $N({\rm{H}_2})$ and $W_{\rm CO}$ for (a) the Polaris Flare, the Polaris molecular cloud, and L1495, and for (b) the Orion A molecular cloud.
\label{fig:NH2_Wco}}
\end{center}
\end{figure}

Figure \ref{fig:NH2_Wco} shows the results.
In the Polaris Flare region (filled black circles), including the Polaris molecular cloud (pink circles), $W_{\rm CO}$ increases linearly with $N({\rm{H}_2})$. 
The conversion factor from $W_{\rm CO}$ to $N$(H$_2$), $X_{\rm CO}$, 
has traditionally been defined as $X_{\rm CO}=W_{\rm CO}$/$N$(H$_2$), assuming a linear relationship between these quantities.
Previous studies have reported a value of $X_{\rm CO}=0.4\times10^{20}$ H$_2$ cm$^{-2}$ (K km s$^{-1})^{-1}$ for the entire Polaris Flare region \citep{Heithausen1990},
while the average value in the solar neighborhood was estimated to be $2.0\times10^{20}$ H$_2$ cm$^{-2}$ (K km s$^{-1})^{-1}$\citep{Strong1996, Dame2001}.
In the present study, we estimated an average $X_{\rm CO}$ value for the Polaris molecular cloud by fitting a linear function forced through the origin to the data shown in Figure \ref{fig:NH2_Wco} (a), although the actual relationship shows some scatter. 
This approach yields a value of $X_{\rm CO}$ as $0.7\times10^{20}$ H$_2$ cm$^{-2}$ (K km s$^{-1})^{-1}$, which is less than half of that in the solar neighborhood.

However, Figure \ref{fig:NH2_Wco} indicates that while the three clouds span different density ranges (Polaris: up to $\sim10^{21}$ cm$^{-2}$, Taurus: $\sim10^{22}$ cm$^{-2}$, and Orion A: $\sim10^{23}$ cm$^{-2}$), 
they show similar trends in their CO-H$_2$ relationships.
The Polaris molecular cloud data points show CO-to-H$_2$ ratios as high as those observed in Taurus at similar densities, despite representing the low-density end of the overall distribution. 
Consequently, while the $X_{\rm CO}$ value of the Polaris molecular cloud is low, it likely falls within the natural diversity observed in the solar neighborhood, albeit at the lower end of the column density distribution.

\begin{figure}
\begin{center}
\includegraphics[scale=.2]{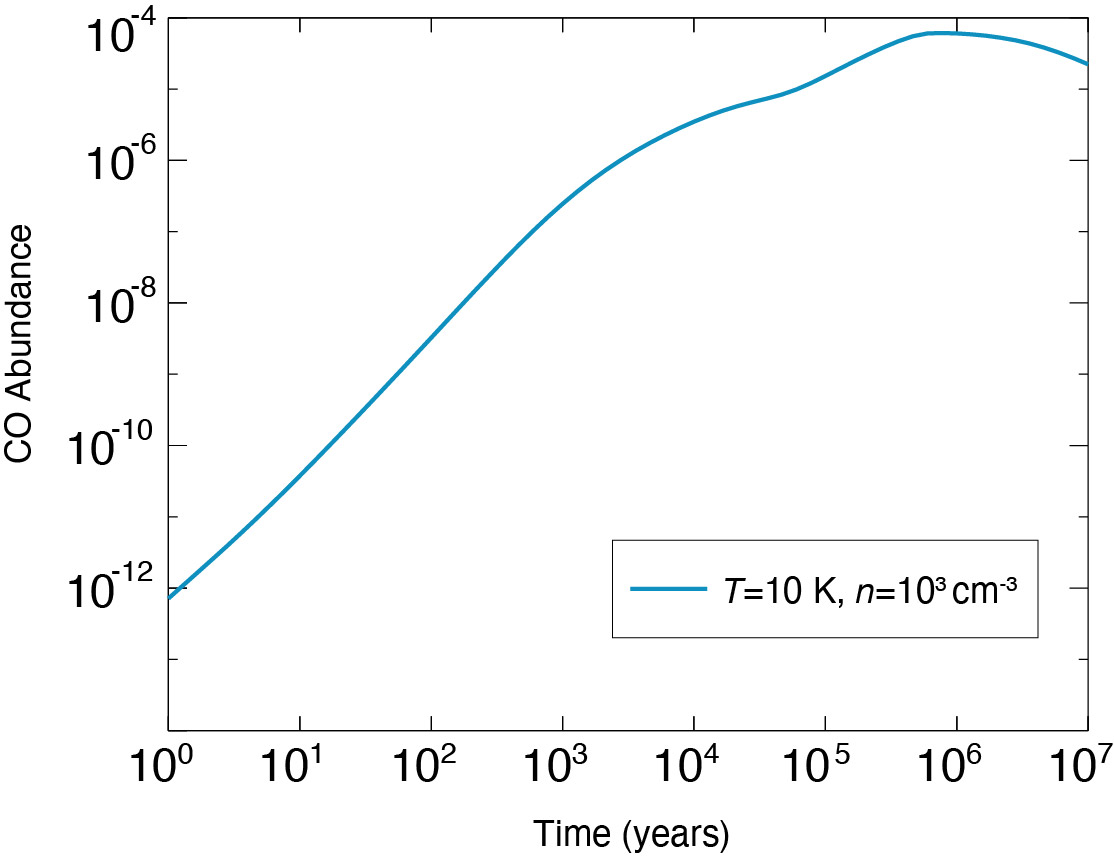}
\caption{
Time evolution of $^{12}$CO abundance from chemical simulations using the Nautilius code  \citep{Ruaud}.
Simulations were performed at a constant temperature of 10 K and gas density of $10^{3}$ cm$^{-3}$, typical for dark clouds. 
The x-axis represents time in years, while the y-axis shows CO abundance relative to H nuclei.
\label{fig:simu}}
\end{center}
\end{figure}

To further investigate the chemical evolution of the Polaris molecular cloud,
we performed chemical simulations of dark cloud models utilizing the astronomical code `Nautilius' \citep{Ruaud}.
It should be noted that our chemical simulations assume that all hydrogen is initially in H$_2$ form, which may not accurately represent the conditions in the Polaris molecular cloud where $\textsc{Hi}$ to H$_2$ conversion is still ongoing.
We applied the same physical parameters and initial abundances as those described by \cite{Taniguchi2019}.
Figure \ref{fig:simu} presents the outcomes of the simulations, tracking the evolution of $^{12}$CO abundance over time. 
The simulations were performed at a constant temperature $T$ of 10 K and a density $n$ of $10^{3}$ cm$^{-3}$
for the gas to simulate conditions typical of dark clouds.
The results demonstrate that the abundance of $^{12}$CO increases along with the time, peaking around $\sim10^{5}-10^{6}$ years before decreasing due to freeze-out onto dust grains.

Assuming $\tau^{{12}\rm{CO}}\ll1$, we attempted to calculate $N(^{12}$CO) using Equation (2), 
and obtained an estimate of the CO fractional abundance relative to $N(\rm{H}_{2})$ to be $\sim1\times10^{-5}$ for the Polaris molecular cloud.
This should be considered as the minimum estimate of the actual fractional abundance, 
because the $^{12}$CO line is not optically thin in reality.
The inferred fractional abundance $\gtrsim1\times10^{-5}$
corresponds to the maximum CO abundance in the simulation, occurring in the time range $10^5-10^6$ years (Figure \ref{fig:simu}).
The simulation indicates a CO formation timescale of $10^5-10^6$ years. 
This relatively high CO abundance is consistent with our finding that the Polaris molecular cloud shows higher CO emission for a given amount of H$_2$ compared to other clouds in the same density range (Figure \ref{fig:NH2_Wco}), and supports the young chemical age of the cloud.
However, this estimate is based on specific initial conditions, including the assumption that all hydrogen is initially in H$_2$ form, which may not accurately represent the conditions in the Polaris molecular cloud.
In the actual Polaris molecular cloud, the atomic hydrogen is still present. 
The CO abundance is determined by the balance between formation and destruction rates of the molecule,
and CO destruction by interstellar UV radiation also could play a role in the diffuse environments. 
In this context, the Polaris molecular cloud and its surrounding Polaris Flare region are characterized by very low densities and the absence of significant star formation, likely resulting in minimal impact from UV radiation on the CO formation and destruction processes.
Despite these factors, 
the estimated time range is consistent with our earlier estimate of $10^5-10^6$ years based on the abundances of early-type molecules such as CCS and HC$_3$N \citep{Shimoikura2012}, 
providing an additional support for the young age of the Polaris molecular cloud.

The lower $X_{\rm CO}$  factor derived for the Polaris molecular cloud, compared to the average in the solar neighborhood, likely reflects the cloud's early stage of chemical evolution. 
This observation aligns with theoretical predictions by \cite{Glover}, who demonstrated through numerical simulations that the CO-to-H$_2$ ratio in molecular clouds can vary significantly depending on factors such as density, metallicity, and radiation field. Their finding that CO formation can be less efficient compared to H$_2$ formation in low-density environments, leading to lower $X_{\rm CO}$ factors, is consistent with our observations of the Polaris molecular cloud.

\subsection{Virial Analysis} 

To understand the overall dynamical state of the molecular cloud, we conducted a virial analysis. Recent theoretical studies suggest that molecular cloud formation and evolution can be driven by global flows in addition to turbulence. Numerical simulations by \cite{Smith} emphasize the role of large-scale flows in the early stages of molecular cloud formation, while \cite{Enrique2019} show that molecular clouds undergo global hierarchical collapse, during which high-density structures are created. The Polaris molecular cloud might also be influenced by both global flows and turbulence.

In preparation for the virial analysis, we determined the surface area $S$ of the cloud using the $^{13}$CO intensity map for the entire observed region (Figure \ref{fig:ii}).
The surface area $S$ was defined as the area where the $^{13}$CO emission exceeds the $3 \,\sigma$ map noise level (=1.2 K km s$^{-1}$).
A Gaussian fit to the average $^{13}$CO spectrum $S$ yielded a line width $\Delta V =\textbf{1.38} $ km s$^{-1}$.
To investigate the potential influence of global flows on the cloud dynamics within $S$, we analyzed $^{13}$CO spectra from three representative 1-pc regions: a dense core region containing cores 4 and 5, an isolated filament region, and a low-density outer region. In smaller regions, we expect turbulent motions to dominate while the influence of global flows should be minimal. 
Therefore, comparing line width in these local regions with that of the entire cloud allows us to evaluate the contribution of organized, global motions.

\begin{figure}
\begin{center}
\includegraphics[scale=0.2]{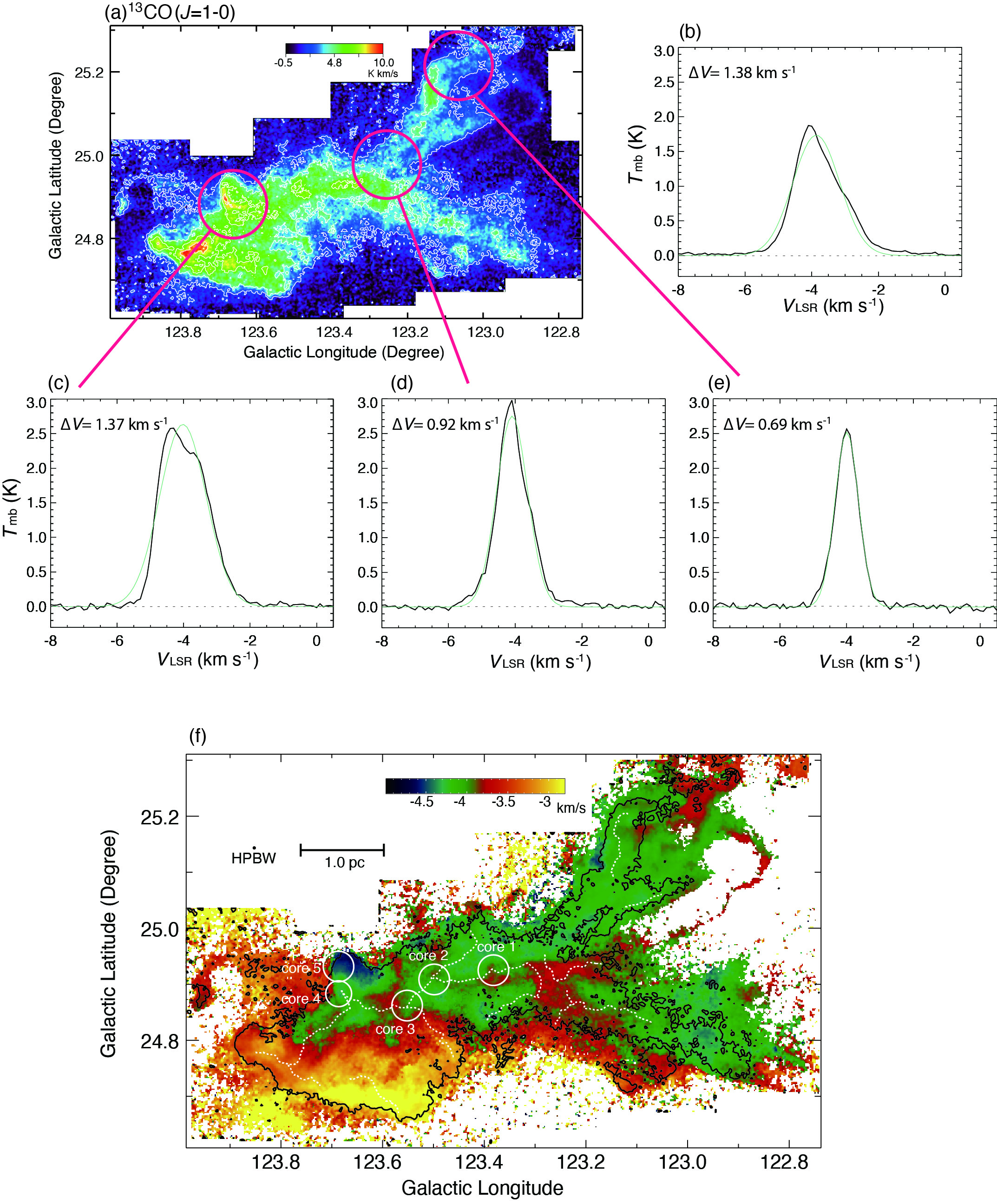}
\caption{
Comparison of $^{13}$CO line profiles in different regions. 
(a) The locations of three selected 1-pc regions (shown as red circles) overlaid on the $^{13}$CO integrated intensity map: a dense core region containing cores 4 and 5, an isolated filament region, and a low-density region. 
(b) Average spectrum of the entire cloud showing a line width of 1.38 km s$^{-1}$. 
(c)--(e) Average spectra within the red circles showing line width of 1.37 km s$^{-1}$, 0.92 km s$^{-1}$, and 0.69 km s$^{-1}$, respectively. 
In panels (b)-(e), the black lines show the observed spectra while the green lines show Gaussian fits.
(f)Distribution of the intensity-weighted mean velocity of the $^{13}$CO emission. 
The black contour represents the 3 K km s$^{-1}$ level of the $^{13}$CO map, 
while the white dotted lines indicate dust filaments identified by HGBS.
The locations of each core (numbered 1-5, same as in Figure \ref{fig:all}) are marked.
\label{fig:V0}}
\end{center}
\end{figure}

Figure \ref{fig:V0}(b)-(e) shows the comparison between the spectrum of area $S$ and those of the local regions.
While $S$ exhibits $\Delta V= 1.38$ km s$^{-1}$, the local regions show systematically different values: 1.37 km s$^{-1}$ in the dense core region, 0.92 km s$^{-1}$ in the filament region, and 0.69 km s$^{-1}$ in the outer region. 
Notably, the line width in the dense region containing cores 4 and 5 is almost identical to that of the entire cloud. 
Given that these dense regions dominate the cloud's mass distribution, and that their line width matches the global value, we conclude that even if global flows exist, their contribution to the observed line width is minimal, having likely been converted into turbulent motions. Therefore, the observed line width can be primarily attributed to turbulent motions.

Based on this understanding, we proceeded with the virial analysis using the line width measured for area $S$.
The radius of the cloud $r$ was calculated as $r=\sqrt{S/\pi}$, yielding a result of $r=2.6$ pc.
The mass of the cloud ($M_{\rm cloud}$) within $S$ was calculated using the following equation,
\begin{equation} 
\label{eq:mass}
{M_{\rm cloud}} = \mu_{\rm H_2} {m_{\rm H}}\int_S {N({\rm{H_2)^{Her}}}ds},
\end{equation}
where $\mu_{\rm H_2}$ is the mean molecular weight assumed to be 2.8 corrected for the $20\%$ helium abundance and $m_{\rm H}$ is the hydrogen mass. 
By integrating $N(\rm{H}_{2})^{Her}$ over the surface area $S$, 
we determined the total cloud mass $M_{\rm cloud}$ to be 310 $M_{\odot}$.


Using these values, we calculated the virial mass $M_{\rm vir}$ using the following equation, 
\begin{equation}
\label{eq:M_Vir}
M_{\rm vir}=
\frac{5r}{G}
\frac{\overline{\Delta V}^{2}}{8\,{\rm ln}2}\ .
\end{equation} 
This calculation yields $M_{\rm vir} = 1066 M_{\odot}$. 
The fact that $M_{\rm vir} > M_{\rm cloud}$ suggests that the Polaris molecular cloud is not in virial equilibrium, but to disperse as a whole.

We further calculated and compared the internal kinetic energy $E_{\rm k}$ and gravitational potential energy $E_{\rm p}$ of the entire cloud using the $^{13}$CO data.

The internal kinetic energy $E_{\rm k}$ is defined as
\begin{equation}
{E_{\rm k}} =\frac{1}{2} M_{\rm cloud} ({\sqrt{3}\, {\sigma}_{v}})^{2},
\end{equation}
where $\sigma_v$ is the velocity dispersion derived from the
$\Delta V$ of \textbf{1.38} km s$^{-1}$ using the relation $\sqrt{8 \rm{ln}2}\, \sigma_v=\Delta V$. 
The factor $\sqrt{3}$ accounts for the three-dimensional velocity dispersion assuming isotropic motions.

The gravitational potential energy $E_{\rm p}$ is defined as
\begin{equation}
{E_{\rm p}} = -\gamma \frac{G M_{\rm cloud}^{2}}{r},
\end{equation}
where $\gamma$ is a factor determined by the shape of the cloud.
We assume $\gamma = \frac{3}{5}$ for a uniform sphere in this paper for simplicity.

Using these parameters, we obtained $E_{\rm k}\approx 160 M_\odot (\rm{km}\, s^{-1})^2$ and $E_{\rm p}\approx -113 M_\odot (\rm{km}\, s^{-1})^2$.
Thus, the quantity $2 E_{\rm k}+E_{\rm p}$ is not equal to 0, but significantly positive ($\approx 200$ $M_\odot (\rm{km}\, s^{-1})^2$), indicating that the kinetic energy substantially exceeds the gravitational potential energy. 

To further investigate the dynamical state of the cloud, we analyzed the velocity structure of the cloud.
We quantified the velocity gradients by investigating the intensity-weighted mean velocity ${V_0}$ map defined as,
\begin{equation}
\label{eq:V0}
V_0=\frac{\int T_{\rm mb}V\,dv}{{\int T_{\rm mb}\,dv}}.
\end{equation}

The calculated ${V_0}$ map for the $^{13}$CO emission line
is presented in Figure \ref{fig:V0}(f), overlaid with the dust filaments identified by the $Herschel$ observations. 
The ${V_0}$ map reveals a prominent large-scale velocity gradient across the molecular cloud,
characterized by a systematic variation in velocity from the northeast to the southwest regions of the cloud. 
This overall gradient spans $\sim 1.5$ km s$^{-1}$ across the entire length of the cloud, indicating a significant large-scale motion or flow within the structure.

\begin{figure}
\begin{center}
\includegraphics[scale=.2]{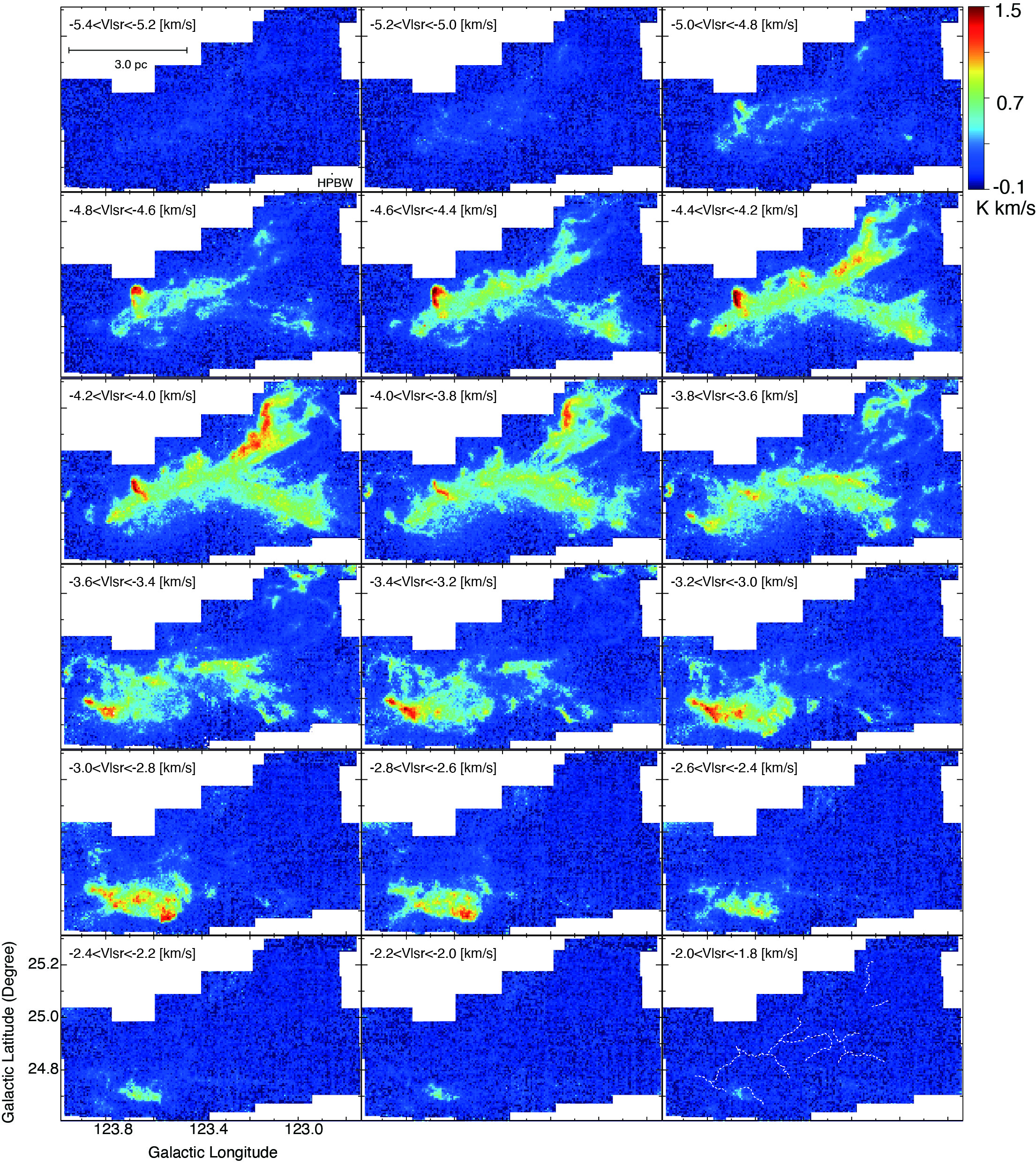}
\caption{
Channel map of the $^{13}$CO emission line created at intervals of 0.2 km s$^{-1}$.
The velocity range for each panel is indicated at the top.
The final panel overlays the distribution of filaments (white dotted lines) identified by HGBS.
\label{fig:channel}}
\end{center}
\end{figure}

To examine the detailed velocity structure along with the dust filaments of the cloud, 
we show the channel map of $^{13}$CO emission in Figure \ref{fig:channel} which reveals
significant variations in the spatial distribution of molecular gas across different velocity channels.

These maps show intricate filamentary structures whose distributions are changing in the velocity channels, indicating the
presence of complex internal motions within the cloud.
Analysis of the velocity structure along filamentary structures seen in Figures \ref{fig:V0} and \ref{fig:channel} reveals systematic velocity gradients ranging from 0.5 to 1.5 km $^{-1}$ pc$^{-1}$.
The range of the velocity gradients are comparable to those found in other molecular clouds \citep[e.g.,][]{Kirk2013}, which are often interpreted as signs of gas accretion along the filaments or global cloud contraction.



\section{Discussion} \label{sec:Discussion}

\begin{figure}
\begin{center}
\includegraphics[scale=.2]{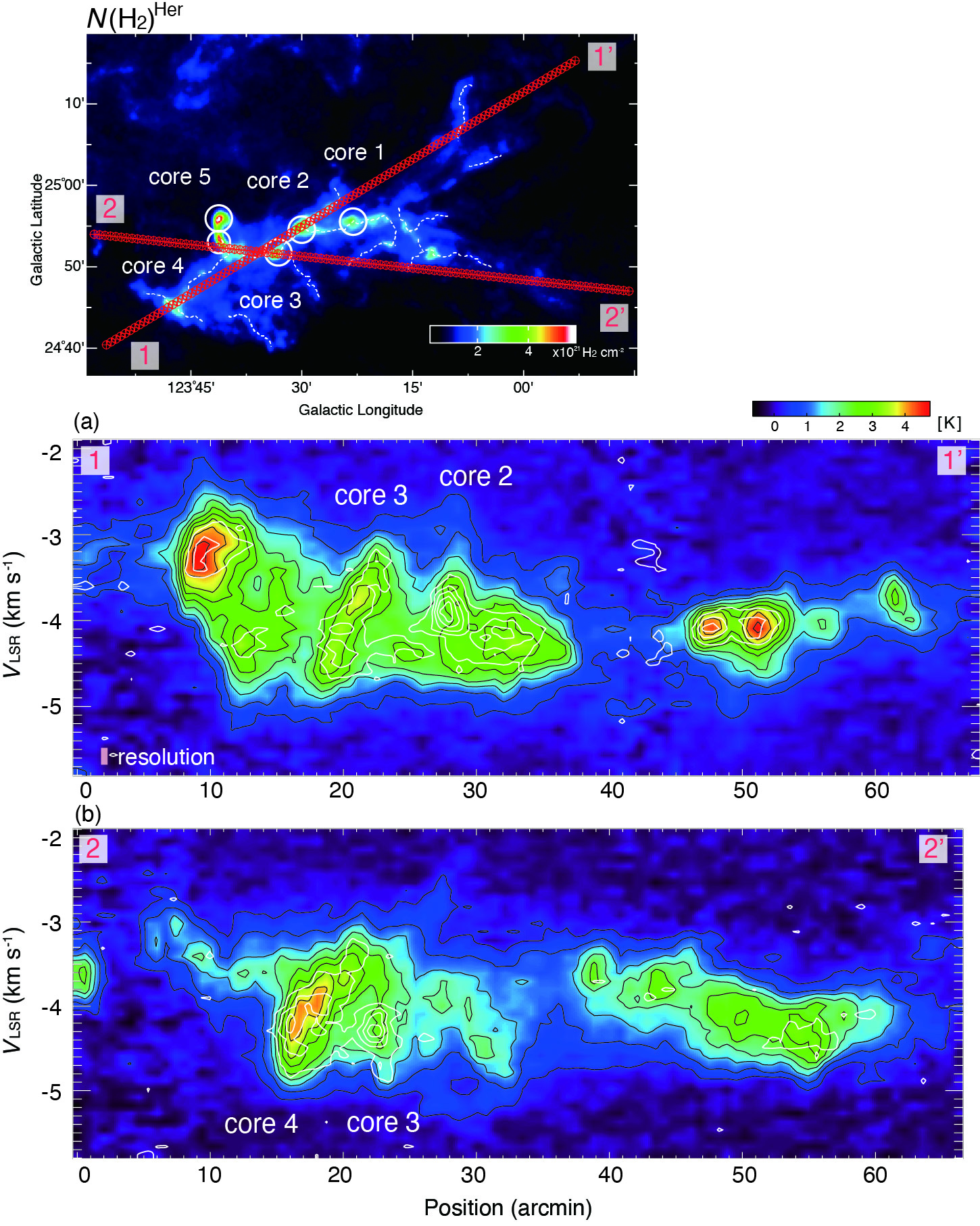}
\caption{
PV diagrams of the $^{13}$CO (color scale and black contour) and C$^{18}$O (white contour) emission lines taken along the cuts labeled 1-1' and 2-2' in the $N(\rm{H}_{2})^{Her}$ map (top panel).
The lowest contours levels are 0.5 K km s$^{-1}$ for $^{13}$CO and 0.3 K km s$^{-1}$ for C$^{18}$O, with the same intervals for both.
\label{fig:PV}}
\end{center}
\end{figure}

The Polaris molecular cloud exhibits distinct velocity structures that reflect the interplay between large-scale dispersion and local gravitational effects.
To investigate this complexity, 
we created Position-Velocity (PV) diagrams for the entire molecular cloud using the $^{13}$CO and C$^{18}$O data, as shown in Figure \ref{fig:PV}.
We divide the Polaris molecular cloud into two parts:
``the main body" containing the five cores at the position $0\arcmin$--$35\arcmin$, 
and ``elongated structures" at the position $>40\arcmin$
in panels (a) and (b) of Figure \ref{fig:PV}.

The velocity structure within the main body exhibits complexity, with diverse internal motions including varying velocity gradients across different cores. 
For instance, cores 2 and 3 in Figure \ref{fig:PV}(a) show velocity gradients in opposite directions, indicating complex internal dynamics.
In contrast, the elongated structures surrounding the main body show simpler velocity gradients. 
These structures appear to be connected to the main body in the PV diagram, with velocities changing smoothly along the structures. 

To examine the velocity structure in more detail, 
we analyzed several elongated structures surrounding the main body, which we refer to as ``sub-filaments".
The sub-filaments, although fainter, are visible around the main body of the cloud in both the $N(\rm{H_2})^{Her}$ map (Figure \ref{fig:all}(b)) and the integrated intensity maps of $^{13}$CO and $^{12}$CO (Figure \ref{fig:ii}).
We selected four sub-filaments with well-defined boundaries and clear connections to the main cloud body for detailed investigation of potential gas accretion processes.

\begin{figure}
\begin{center}
\includegraphics[scale=0.2]{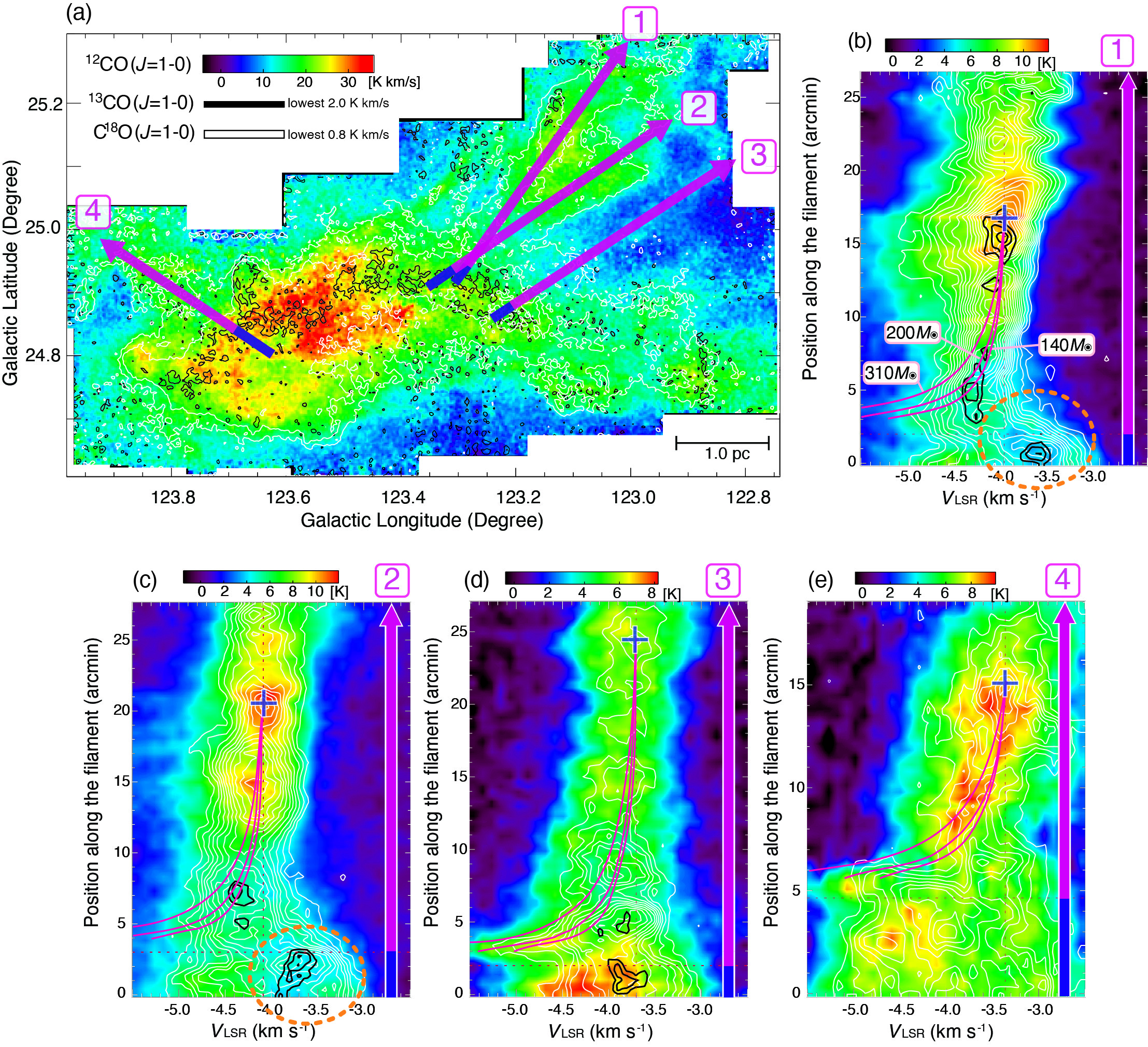}
\caption{
Analysis of the structure and gas motion of the Polaris molecular cloud. 
(a) Integrated intensity map of the Polaris molecular cloud. 
The color map represents the $^{12}$CO emission line, with the white contours showing the $^{13}$CO emission line at 2.0 K km s$^{-1}$ and 4.0 K km s$^{-1}$, 
and the black contours showing the C$^{18}$O emission line at 0.8 K km s$^{-1}$ and 1.0 K km s$^{-1}$. 
Arrows indicate the positions of the PV diagrams along selected sub-filaments. 
(b)-(e) PV diagrams of each sub-filament. 
The color map represents $^{12}$CO, white contours represent $^{13}$CO, and black contours represent C$^{18}$O, as in panel (a).
The $^{13}$CO contours are drawn at intervals of 0.3 K, starting from 0.6 K, while the C$^{18}$O contours are drawn at intervals of 0.2 K, starting from 0.6 K. 
The three curves represent free-fall models for different masses of the main body: $140\,M_{\odot}$, $200\,M_{\odot}$, and $310\,M_{\odot}$. 
The `+' symbols mark the locations of gas motion measurements.
\label{fig:filamentPV}}
\end{center}
\end{figure}

Figures \ref{fig:filamentPV}(b)-(e) show high-resolution PV diagrams of the selected sub-filaments
with $^{12}$CO (color), $^{13}$CO (white contours), and C$^{18}$O (black contours).
The sub-filaments appear to be connected to the main body, with velocities changing smoothly toward the lower left direction. This pattern suggests that the gas within these sub-filaments is accelerating and falling toward the main body, providing direct evidence for the accretion of material onto the main cloud structure.

To analyze the gas motion quantitatively, we defined reference points for each sub-filament labeled 1-4 in Figure \ref{fig:filamentPV}(a) as the C$^{18}$O emission peak within the region enclosed by the $4\,\sigma$ contour of C$^{18}$O emission where the sub-filament connects to the main body. 
These reference points, considered as the destinations of infalling gas, are indicated by the boundaries between the purple and blue sections of the arrows in Figure \ref{fig:filamentPV}(b)-(e). 
Along each sub-filament, we selected measurement positions (marked by `+' symbols) where clear emission is detected in both $^{13}$CO and $^{12}$CO. 
We then measured both the projected distances from these positions to their respective reference points and their corresponding $V_{\rm LSR}$ values.

We tested the free-fall hypothesis for the sub-filaments by estimating the free-fall velocity as a function of the distance to the main body. 
Assuming that the gas in the sub-filaments is free-falling towards the main body, 
we estimated the gas velocity $v$ and its distance $x$ from the main body at positions along the sub-filament using the following equation
\begin{equation}
|v(x)|= \sqrt{v_{0}^2+2GM\left({\frac{1}{x}-{\frac{1}{x_{0}}}}\right)},
\end{equation}
where $M$ is the mass of the main body of the cloud, $v_{0}$ is the observed velocity at $x=x_{0}$. 

We measured the values of $v_{0}$ and $x_{0}$ at the positions marked by the `+'
signs along the sub-filaments shown in Figure \ref{fig:filamentPV}(b)-(e). 
Due to various uncertainties in the cloud mass, we calculated $v$ for three different
masses:
$140\,M_{\odot}$ (total mass contained in regions with $N(\rm{H}_{2})^{Her}>1\times10^{21} cm^{-2}$, see Figure \ref{fig:all}(b)),
$200\,M_{\odot}$ (total mass contained in regions with $^{13}$CO integrated intensity $> 2.0$ K km s$^{-1}$, or $5\,\sigma$ of the map noise level), and
$310\,M_{\odot}$ (total mass contained in regions with $^{13}$CO integrated intensity $> 1.2$ K km s$^{-1}$, or $3\, \sigma$ of the map noise level).

The resulting $v(x)$ values are plotted in each PV diagram by the red lines. 
These $v(x)$ values show an agreement with the observed velocity gradients within the range of the assumed masses, though we disregarded the inclination angles of the sub-filaments to the line of sight of the observers. 
These comparisons support the free-fall hypothesis.

In the observed PV diagrams, opposite velocity gradients are seen around the surface of the main body (outlined by orange dashed lines in panels (b) and (c) of Figure \ref{fig:filamentPV}). 
The velocity gradient changes its sign at the boundary of the main body, showing an opposite direction to that of the infalling gas. 
This reversal of the velocity gradient suggests the deceleration and possible accumulation of the infalling gas as it encounters the dense gas of the main body, similar to a shock-like interface between the infalling and the ambient gas.

The above analysis infers that the gas in the subfilaments is likely free-falling toward the main body of the Polaris molecular cloud, leading to mass accumulation to the molecular cloud from the surroundings. 
This hub-filament morphology and associated velocity structure is consistent with patterns seen in numerical simulations of cloud formation \citep{Gomez2014} and observations of more evolved molecular clouds \citep{Hacar2017}. 
The velocity patterns we observe in the Polaris cloud's filaments, particularly the systematic velocity gradients pointing toward the main cloud body as shown in Figure \ref{fig:filamentPV}, are remarkably similar to the converging flows predicted by gravitational collapse models of hub-filament systems \citep[e.g., Figure 5 in][]{Gomez2014}. However, our observations capture this process at an earlier evolutionary stage and lower column density than previously observed, providing new insights into the initial phases of hub-filament formation.
Similar infalling motion along subfilaments have been found in some other molecular clouds
\citep[e.g.,][]{Kirk2013,her1,Chen2019,Shimoikura2016,Shimoikura2022},
but it is noteworthy that they are much denser ($10^{22} - 10^{23}$ cm$^{-2}$)
compared to the Polaris molecular clouds. 

\cite{Shimoikura2012} found virial ratios of approximately 2 for the cores in the Polaris molecular cloud (ie., cores 4 and 5 or MCLD123.5+24.9), suggesting that the cores can be close to the virial equilibrium state. 
This previous finding, combined with our detection of infall motions along the sub-filaments, presents an interesting picture of the cloud's evolution. While the cloud as a whole remains gravitationally unbound, the continuous mass accretion through the sub-filaments appears to be building up dense cores that approach virial equilibrium.
This suggests that local gravitational effects, enhanced by the mass accumulation along filaments, play a crucial role in creating and maintaining bound structures even in low-density environments.

This study demonstrates that gravitational effects can be significant even at column densities of $\sim 10^{21}$ cm$^{-2}$, much lower than previous detections in other clouds ($10^{22}-10^{23}$ cm$^{-2}$).
While our virial analysis indicates that the entire cloud is gravitationally unbound, suggesting that the cloud evolution is not primarily driven by gravity at this stage, we detect clear signatures of gravitational influence in the filamentary structures. The observed free-fall patterns in the extended filaments indicate that the cloud is in a transition phase where gravity is beginning to play an important role. This unique state makes the Polaris molecular cloud an ideal laboratory for studying the early stages of gravitationally-influenced cloud evolution.

The observations presented here reveal an interaction between global and local dynamics in the Polaris molecular cloud, indicating the multi-scale nature of cloud dynamics. This finding aligns with theoretical predictions from hierarchical cloud formation models \citep{Smith} , where large-scale flows create conditions for localized gravitational collapse, and global hierarchical collapse scenarios \citep{Enrique2019} where gravity drives cloud evolution across multiple scales simultaneously. The Polaris molecular cloud's location at high galactic latitude makes it particularly suitable for testing these theoretical models, as it is relatively isolated from external perturbations that could complicate the dynamics in galactic plane clouds.

\section{Summary} \label{sec:conclusions}

We performed extensive mapping observations of the Polaris molecular cloud, a dense region within the Polaris Flare, utilizing the Nobeyama 45 m Radio Telescope to observe $J=1-0$ line of $^{12}$CO, $^{13}$CO, and C$^{18}$O.
Our main findings are as follows:

\begin{enumerate}
\item
The observations reveal molecular gas formation at column densities up to $\sim10^{21}$ cm$^{-2}$, evidenced by the anti-correlation between $\textsc{Hi}$ and CO distributions. 
We found a threshold column density for molecular formation at $\sim 5\times10^{20}$ cm$^{-2}$, which is common among more evolved molecular clouds.
The presence of $\textsc{Hi}$ gas and its anti-correlation with CO emission indicates ongoing conversion from atomic to molecular gas.

\item
The CO-to-H$_2$ conversion factor, $X_{\rm CO}$=$0.7\times10^{20}$ H$_2$ cm$^{-2}$ (K km s$^{-1})^{-1}$, is lower than the solar neighborhood average of $2.0\times10^{20}$ H$_2$ cm$^{-2}$ (K km s$^{-1})^{-1}$.
At comparable densities, however, the Polaris molecular cloud shows CO emission as high as that observed in more evolved clouds like Taurus.
Chemical models suggest a cloud age of $\sim10^{5}-10^{6}$ years, indicating an early stage of molecular cloud evolution.

\item
The virial analysis indicates that while the cloud is gravitationally unbound at this stage, we identified several elongated structures extending from the main cloud body showing systematic velocity gradients of $0.5-1.5$ km s$^{-1}$ pc$^{-1}$. 
The observed velocities in these structures are consistent with free-fall models, and show signs of deceleration at the main cloud surface, providing evidence for gravitational infall and mass accretion toward the main cloud body. 
This mass accumulation process, occurring simultaneously with the molecular gas formation, demonstrates that gravitational effects can be significant even in relatively low-density environments ($\sim10^{21}$ cm$^{-2}$). 
This makes the Polaris molecular cloud an ideal target for studying the early stages of molecular cloud evolution.
\end{enumerate}


\begin{acknowledgments}

We are very grateful to the anonymous referee for providing useful comments and suggestions to improve this paper.
This work was supported by JSPS KAKENHI (Nos. 20K14523, 21H01142, 22K02966, 24H0025, 24K00463, 24K17096, 22K18618). 
T.S. also acknowledges support from the Kayamori Foundation of Informational Science Advancement (K33 ken XXXVI591).
The Nobeyama 45-m radio telescope is operated by Nobeyama Radio Observatory, a branch of National Astronomical Observatory of Japan.

\end{acknowledgments}


\begin{thebibliography}{}
\expandafter\ifx\csname natexlab\endcsname\relax\def\natexlab#1{#1}\fi
\providecommand{\url}[1]{\href{#1}{#1}}
\providecommand{\dodoi}[1]{doi:~\href{http://doi.org/#1}{\nolinkurl{#1}}}
\providecommand{\doeprint}[1]{\href{http://ascl.net/#1}{\nolinkurl{http://ascl.net/#1}}}
\providecommand{\doarXiv}[1]{\href{https://arxiv.org/abs/#1}{\nolinkurl{https://arxiv.org/abs/#1}}}

\bibitem[{{Andr{\'e}} {et~al.}(2013){Andr{\'e}}, {K{\"o}nyves}, {Arzoumanian},
  {Palmeirim}, \& {Peretto}}]{Andre2013}
{Andr{\'e}}, P., {K{\"o}nyves}, V., {Arzoumanian}, D., {Palmeirim}, P., \&
  {Peretto}, N. 2013, in Astronomical Society of the Pacific Conference Series,
  Vol. 476, New Trends in Radio Astronomy in the ALMA Era: The 30th Anniversary
  of Nobeyama Radio Observatory, ed. R.~{Kawabe}, N.~{Kuno}, \& S.~{Yamamoto},
  95

\bibitem[{{Andr{\'e}} {et~al.}(2010){Andr{\'e}}, {Men'shchikov}, {Bontemps},
  {K{\"o}nyves}, {Motte}, {Schneider}, {Didelon}, {Minier}, {Saraceno},
  {Ward-Thompson}, {di Francesco}, {White}, {Molinari}, {Testi}, {Abergel},
  {Griffin}, {Henning}, {Royer}, {Mer{\'\i}n}, {Vavrek}, {Attard},
  {Arzoumanian}, {Wilson}, {Ade}, {Aussel}, {Baluteau}, {Benedettini},
  {Bernard}, {Blommaert}, {Cambr{\'e}sy}, {Cox}, {di Giorgio}, {Hargrave},
  {Hennemann}, {Huang}, {Kirk}, {Krause}, {Launhardt}, {Leeks}, {Le Pennec},
  {Li}, {Martin}, {Maury}, {Olofsson}, {Omont}, {Peretto}, {Pezzuto}, {Prusti},
  {Roussel}, {Russeil}, {Sauvage}, {Sibthorpe}, {Sicilia-Aguilar}, {Spinoglio},
  {Waelkens}, {Woodcraft}, \& {Zavagno}}]{Andre2010}
{Andr{\'e}}, P., {Men'shchikov}, A., {Bontemps}, S., {et~al.} 2010, \aap, 518,
  L102, \dodoi{10.1051/0004-6361/201014666}

\bibitem[{{Arzoumanian} {et~al.}(2019){Arzoumanian}, {Andr{\'e}},
  {K{\"o}nyves}, {Palmeirim}, {Roy}, {Schneider}, {Benedettini}, {Didelon}, {Di
  Francesco}, {Kirk}, \& {Ladjelate}}]{Arzoumanian2019}
{Arzoumanian}, D., {Andr{\'e}}, P., {K{\"o}nyves}, V., {et~al.} 2019, \aap,
  621, A42, \dodoi{10.1051/0004-6361/201832725}

\bibitem[{{Bernard} {et~al.}(1999){Bernard}, {Abergel}, {Ristorcelli}, {Pajot},
  {Torre}, {Boulanger}, {Giard}, {Lagache}, {Serra}, {Lamarre}, {Puget},
  {Lepeintre}, \& {Cambr{\'e}sy}}]{Bernard1999}
{Bernard}, J.~P., {Abergel}, A., {Ristorcelli}, I., {et~al.} 1999, \aap, 347,
  640

\bibitem[{{Chen} {et~al.}(2019){Chen}, {Zhang}, {Wright}, {Busquet}, {Lin},
  {Liu}, {Olguin}, {Sanhueza}, {Nakamura}, {Palau}, {Ohashi}, {Tatematsu}, \&
  {Liao}}]{Chen2019}
{Chen}, H.-R.~V., {Zhang}, Q., {Wright}, M.~C.~H., {et~al.} 2019, \apj, 875,
  24, \dodoi{10.3847/1538-4357/ab0f3e}

\bibitem[{{Dame} {et~al.}(2001){Dame}, {Hartmann}, \& {Thaddeus}}]{Dame2001}
{Dame}, T.~M., {Hartmann}, D., \& {Thaddeus}, P. 2001, \apj, 547, 792,
  \dodoi{10.1086/318388}

\bibitem[{{Dobashi}(2011)}]{Dobashi2011}
{Dobashi}, K. 2011, \pasj, 63, 1, \dodoi{10.1093/pasj/63.sp1.S1}

\bibitem[{{Dobashi} {et~al.}(2005){Dobashi}, {Uehara}, {Kandori}, {Sakurai},
  {Kaiden}, {Umemoto}, \& {Sato}}]{Dobashi2005}
{Dobashi}, K., {Uehara}, H., {Kandori}, R., {et~al.} 2005, \pasj, 57, 1

\bibitem[{{Glover} \& {Mac Low}(2011)}]{Glover}
{Glover}, S.~C.~O., \& {Mac Low}, M.~M. 2011, \mnras, 412, 337,
  \dodoi{10.1111/j.1365-2966.2010.17907.x}

\bibitem[{{G{\'o}mez} \& {V{\'a}zquez-Semadeni}(2014)}]{Gomez2014}
{G{\'o}mez}, G.~C., \& {V{\'a}zquez-Semadeni}, E. 2014, \apj, 791, 124,
  \dodoi{10.1088/0004-637X/791/2/124}

\bibitem[{{Grossmann} \& {Heithausen}(1992)}]{Grossmann1992}
{Grossmann}, V., \& {Heithausen}, A. 1992, \aap, 264, 195

\bibitem[{{Hacar} {et~al.}(2016){Hacar}, {Alves}, {Burkert}, \&
  {Goldsmith}}]{Hacar2016}
{Hacar}, A., {Alves}, J., {Burkert}, A., \& {Goldsmith}, P. 2016, \aap, 591,
  A104, \dodoi{10.1051/0004-6361/201527319}

\bibitem[{{Hacar} {et~al.}(2017){Hacar}, {Tafalla}, \& {Alves}}]{Hacar2017}
{Hacar}, A., {Tafalla}, M., \& {Alves}, J. 2017, \aap, 606, A123,
  \dodoi{10.1051/0004-6361/201630348}

\bibitem[{{Hearty}(1999)}]{Hearty1999}
{Hearty}, T. 1999, in Japanese-German Workshop on High Energy Astrophysics, ed.
  W.~{Becker} \& M.~{Itoh}, 11

\bibitem[{{Heithausen}(1999)}]{Heithausen1999}
{Heithausen}, A. 1999, \aap, 349, L53

\bibitem[{{Heithausen} {et~al.}(1993){Heithausen}, {Stacy}, {de Vries},
  {Mebold}, \& {Thaddeus}}]{Heithausen1993}
{Heithausen}, A., {Stacy}, J.~G., {de Vries}, H.~W., {Mebold}, U., \&
  {Thaddeus}, P. 1993, \aap, 268, 265

\bibitem[{{Heithausen} \& {Thaddeus}(1990)}]{Heithausen1990}
{Heithausen}, A., \& {Thaddeus}, P. 1990, \apjl, 353, L49,
  \dodoi{10.1086/185705}

\bibitem[{{Heitsch} {et~al.}(2009){Heitsch}, {Ballesteros-Paredes}, \&
  {Hartmann}}]{Heitsch}
{Heitsch}, F., {Ballesteros-Paredes}, J., \& {Hartmann}, L. 2009, \apj, 704,
  1735, \dodoi{10.1088/0004-637X/704/2/1735}

\bibitem[{{HI4PI Collaboration} {et~al.}(2016){HI4PI Collaboration}, {Ben
  Bekhti}, {Fl{\"o}er}, {Keller}, {Kerp}, {Lenz}, {Winkel}, {Bailin},
  {Calabretta}, {Dedes}, {Ford}, {Gibson}, {Haud}, {Janowiecki}, {Kalberla},
  {Lockman}, {McClure-Griffiths}, {Murphy}, {Nakanishi}, {Pisano}, \&
  {Staveley-Smith}}]{HI4PI}
{HI4PI Collaboration}, {Ben Bekhti}, N., {Fl{\"o}er}, L., {et~al.} 2016, \aap,
  594, A116, \dodoi{10.1051/0004-6361/201629178}

\bibitem[{{Kirk} {et~al.}(2013{\natexlab{a}}){Kirk}, {Myers}, {Bourke},
  {Gutermuth}, {Hedden}, \& {Wilson}}]{Kirk2013}
{Kirk}, H., {Myers}, P.~C., {Bourke}, T.~L., {et~al.} 2013{\natexlab{a}}, \apj,
  766, 115, \dodoi{10.1088/0004-637X/766/2/115}

\bibitem[{{Kirk} {et~al.}(2013{\natexlab{b}}){Kirk}, {Ward-Thompson},
  {Palmeirim}, {Andr{\'e}}, {Griffin}, {Hargrave}, {K{\"o}nyves}, {Bernard},
  {Nutter}, {Sibthorpe}, {Di Francesco}, {Abergel}, {Arzoumanian},
  {Benedettini}, {Bontemps}, {Elia}, {Hennemann}, {Hill}, {Men'shchikov},
  {Motte}, {Nguyen-Luong}, {Peretto}, {Pezzuto}, {Rygl}, {Sadavoy}, {Schisano},
  {Schneider}, {Testi}, \& {White}}]{her2}
{Kirk}, J.~M., {Ward-Thompson}, D., {Palmeirim}, P., {et~al.}
  2013{\natexlab{b}}, \mnras, 432, 1424, \dodoi{10.1093/mnras/stt561}

\bibitem[{{Lovas}(1992)}]{Lovas}
{Lovas}, F.~J. 1992, Journal of Physical and Chemical Reference Data, 21, 181,
  \dodoi{10.1063/1.555920}

\bibitem[{{Magnani} {et~al.}(1985){Magnani}, {Blitz}, \& {Mundy}}]{Magnani1985}
{Magnani}, L., {Blitz}, L., \& {Mundy}, L. 1985, \apj, 295, 402,
  \dodoi{10.1086/163385}

\bibitem[{{Mangum} \& {Shirley}(2015)}]{Mangum2015}
{Mangum}, J.~G., \& {Shirley}, Y.~L. 2015, \pasp, 127, 266,
  \dodoi{10.1086/680323}

\bibitem[{{Marsh} {et~al.}(2016){Marsh}, {Kirk}, {Andr{\'e}}, {Griffin},
  {K{\"o}nyves}, {Palmeirim}, {Men'shchikov}, {Ward-Thompson}, {Benedettini},
  {Bresnahan}, {di Francesco}, {Elia}, {Motte}, {Peretto}, {Pezzuto}, {Roy},
  {Sadavoy}, {Schneider}, {Spinoglio}, \& {White}}]{her3}
{Marsh}, K.~A., {Kirk}, J.~M., {Andr{\'e}}, P., {et~al.} 2016, \mnras, 459,
  342, \dodoi{10.1093/mnras/stw301}

\bibitem[{{Men'shchikov} {et~al.}(2010){Men'shchikov}, {Andr{\'e}}, {Didelon},
  {K{\"o}nyves}, {Schneider}, {Motte}, {Bontemps}, {Arzoumanian}, {Attard},
  {Abergel}, {Baluteau}, {Bernard}, {Cambr{\'e}sy}, {Cox}, {di Francesco}, {di
  Giorgio}, {Griffin}, {Hargrave}, {Huang}, {Kirk}, {Li}, {Martin}, {Minier},
  {Miville-Desch{\^e}nes}, {Molinari}, {Olofsson}, {Pezzuto}, {Roussel},
  {Russeil}, {Saraceno}, {Sauvage}, {Sibthorpe}, {Spinoglio}, {Testi},
  {Ward-Thompson}, {White}, {Wilson}, {Woodcraft}, \&
  {Zavagno}}]{Menshchikov2010}
{Men'shchikov}, A., {Andr{\'e}}, P., {Didelon}, P., {et~al.} 2010, \aap, 518,
  L103, \dodoi{10.1051/0004-6361/201014668}

\bibitem[{{Minamidani} {et~al.}(2016){Minamidani}, {Nishimura}, {Miyamoto},
  {Kaneko}, {Iwashita}, {Miyazawa}, {Nishitani}, {Wada}, {Fujii}, {Takahashi},
  {Iizuka}, {Ogawa}, {Kimura}, {Kozuki}, {Hasegawa}, {Matsuo}, {Fujita},
  {Ohashi}, {Morokuma-Matsui}, {Maekawa}, {Muraoka}, {Nakajima}, {Umemoto},
  {Sorai}, {Nakamura}, {Kuno}, \& {Saito}}]{Minamidani}
{Minamidani}, T., {Nishimura}, A., {Miyamoto}, Y., {et~al.} 2016, in \procspie,
  Vol. 9914, Millimeter, Submillimeter, and Far-Infrared Detectors and
  Instrumentation for Astronomy VIII, 99141Z, \dodoi{10.1117/12.2232137}

\bibitem[{{Miville-Desch{\^e}nes} {et~al.}(2010){Miville-Desch{\^e}nes},
  {Martin}, {Abergel}, {Bernard}, {Boulanger}, {Lagache}, {Anderson},
  {Andr{\'e}}, {Arab}, {Baluteau}, {Blagrave}, {Bontemps}, {Cohen},
  {Compiegne}, {Cox}, {Dartois}, {Davis}, {Emery}, {Fulton}, {Gry}, {Habart},
  {Huang}, {Joblin}, {Jones}, {Kirk}, {Lim}, {Madden}, {Makiwa}, {Menshchikov},
  {Molinari}, {Moseley}, {Motte}, {Naylor}, {Okumura}, {Pinheiro
  Gon{\c{c}}alves}, {Polehampton}, {Rod{\'o}n}, {Russeil}, {Saraceno},
  {Schneider}, {Sidher}, {Spencer}, {Swinyard}, {Ward-Thompson}, {White}, \&
  {Zavagno}}]{Miville2010}
{Miville-Desch{\^e}nes}, M.~A., {Martin}, P.~G., {Abergel}, A., {et~al.} 2010,
  \aap, 518, L104, \dodoi{10.1051/0004-6361/201014678}

\bibitem[{{Palmeirim} {et~al.}(2013){Palmeirim}, {Andr{\'e}}, {Kirk},
  {Ward-Thompson}, {Arzoumanian}, {K{\"o}nyves}, {Didelon}, {Schneider},
  {Benedettini}, {Bontemps}, {Di Francesco}, {Elia}, {Griffin}, {Hennemann},
  {Hill}, {Martin}, {Men'shchikov}, {Molinari}, {Motte}, {Nguyen Luong},
  {Nutter}, {Peretto}, {Pezzuto}, {Roy}, {Rygl}, {Spinoglio}, \&
  {White}}]{her1}
{Palmeirim}, P., {Andr{\'e}}, P., {Kirk}, J., {et~al.} 2013, \aap, 550, A38,
  \dodoi{10.1051/0004-6361/201220500}

\bibitem[{{Panopoulou} {et~al.}(2022){Panopoulou}, {Clark}, {Hacar}, {Heitsch},
  {Kainulainen}, {Ntormousi}, {Seifried}, \& {Smith}}]{Panopoulou}
{Panopoulou}, G.~V., {Clark}, S.~E., {Hacar}, A., {et~al.} 2022, \aap, 657,
  L13, \dodoi{10.1051/0004-6361/202142281}

\bibitem[{{Polychroni} {et~al.}(2013){Polychroni}, {Schisano}, {Elia}, {Roy},
  {Molinari}, {Martin}, {Andr{\'e}}, {Turrini}, {Rygl}, {Di Francesco},
  {Benedettini}, {Busquet}, {di Giorgio}, {Pestalozzi}, {Pezzuto},
  {Arzoumanian}, {Bontemps}, {Hennemann}, {Hill}, {K{\"o}nyves},
  {Men'shchikov}, {Motte}, {Nguyen-Luong}, {Peretto}, {Schneider}, \&
  {White}}]{her5}
{Polychroni}, D., {Schisano}, E., {Elia}, D., {et~al.} 2013, \apjl, 777, L33,
  \dodoi{10.1088/2041-8205/777/2/L33}

\bibitem[{{Roy} {et~al.}(2013){Roy}, {Martin}, {Polychroni}, {Bontemps},
  {Abergel}, {Andr{\'e}}, {Arzoumanian}, {Di Francesco}, {Hill}, {Konyves},
  {Nguyen-Luong}, {Pezzuto}, {Schneider}, {Testi}, \& {White}}]{her4}
{Roy}, A., {Martin}, P.~G., {Polychroni}, D., {et~al.} 2013, \apj, 763, 55,
  \dodoi{10.1088/0004-637X/763/1/55}

\bibitem[{{Ruaud} {et~al.}(2016){Ruaud}, {Wakelam}, \& {Hersant}}]{Ruaud}
{Ruaud}, M., {Wakelam}, V., \& {Hersant}, F. 2016, \mnras, 459, 3756,
  \dodoi{10.1093/mnras/stw887}

\bibitem[{{Sawada} {et~al.}(2008){Sawada}, {Ikeda}, {Sunada}, {Kuno},
  {Kamazaki}, {Morita}, {Kurono}, {Koura}, {Abe}, {Kawase}, {Maekawa},
  {Horigome}, \& {Yanagisawa}}]{Sawada}
{Sawada}, T., {Ikeda}, N., {Sunada}, K., {et~al.} 2008, \pasj, 60, 445,
  \dodoi{10.1093/pasj/60.3.445}

\bibitem[{{Shimoikura} {et~al.}(2022){Shimoikura}, {Dobashi}, {Hirano},
  {Nakamura}, {Hirota}, {Matsumoto}, {Taniguchi}, \&
  {Shimajiri}}]{Shimoikura2022}
{Shimoikura}, T., {Dobashi}, K., {Hirano}, N., {et~al.} 2022, \apj, 928, 76,
  \dodoi{10.3847/1538-4357/ac5327}

\bibitem[{{Shimoikura} {et~al.}(2016){Shimoikura}, {Dobashi}, {Matsumoto}, \&
  {Nakamura}}]{Shimoikura2016}
{Shimoikura}, T., {Dobashi}, K., {Matsumoto}, T., \& {Nakamura}, F. 2016, \apj,
  832, 205, \dodoi{10.3847/0004-637X/832/2/205}

\bibitem[{{Shimoikura} {et~al.}(2012){Shimoikura}, {Dobashi}, {Sakurai},
  {Takano}, {Nishiura}, \& {Hirota}}]{Shimoikura2012}
{Shimoikura}, T., {Dobashi}, K., {Sakurai}, T., {et~al.} 2012, \apj, 745, 195,
  \dodoi{10.1088/0004-637X/745/2/195}

\bibitem[{{Smith} {et~al.}(2016){Smith}, {Glover}, {Klessen}, \&
  {Fuller}}]{Smith}
{Smith}, R.~J., {Glover}, S. C.~O., {Klessen}, R.~S., \& {Fuller}, G.~A. 2016,
  \mnras, 455, 3640, \dodoi{10.1093/mnras/stv2559}

\bibitem[{{Strong} \& {Mattox}(1996)}]{Strong1996}
{Strong}, A.~W., \& {Mattox}, J.~R. 1996, \aap, 308, L21

\bibitem[{{Suzuki} {et~al.}(1992){Suzuki}, {Yamamoto}, {Ohishi}, {Kaifu},
  {Ishikawa}, {Hirahara}, \& {Takano}}]{Suzuki1992}
{Suzuki}, H., {Yamamoto}, S., {Ohishi}, M., {et~al.} 1992, \apj, 392, 551,
  \dodoi{10.1086/171456}

\bibitem[{{Taniguchi} {et~al.}(2019){Taniguchi}, {Herbst}, {Caselli},
  {Paulive}, {Maffucci}, \& {Saito}}]{Taniguchi2019}
{Taniguchi}, K., {Herbst}, E., {Caselli}, P., {et~al.} 2019, \apj, 881, 57,
  \dodoi{10.3847/1538-4357/ab2d9e}

\bibitem[{{V{\'a}zquez-Semadeni} {et~al.}(2019){V{\'a}zquez-Semadeni}, {Palau},
  {Ballesteros-Paredes}, {G{\'o}mez}, \& {Zamora-Avil{\'e}s}}]{Enrique2019}
{V{\'a}zquez-Semadeni}, E., {Palau}, A., {Ballesteros-Paredes}, J.,
  {G{\'o}mez}, G.~C., \& {Zamora-Avil{\'e}s}, M. 2019, \mnras, 490, 3061,
  \dodoi{10.1093/mnras/stz2736}

\bibitem[{{Ward-Thompson} {et~al.}(2010){Ward-Thompson}, {Kirk}, {Andr{\'e}},
  {Saraceno}, {Didelon}, {K{\"o}nyves}, {Schneider}, {Abergel}, {Baluteau},
  {Bernard}, {Bontemps}, {Cambr{\'e}sy}, {Cox}, {di Francesco}, {di Giorgio},
  {Griffin}, {Hargrave}, {Huang}, {Li}, {Martin}, {Men'shchikov}, {Minier},
  {Molinari}, {Motte}, {Olofsson}, {Pezzuto}, {Russeil}, {Sauvage},
  {Sibthorpe}, {Spinoglio}, {Testi}, {White}, {Wilson}, {Woodcraft}, \&
  {Zavagno}}]{Ward2010}
{Ward-Thompson}, D., {Kirk}, J.~M., {Andr{\'e}}, P., {et~al.} 2010, \aap, 518,
  L92, \dodoi{10.1051/0004-6361/201014618}

\bibitem[{{Zagury} {et~al.}(1999){Zagury}, {Boulanger}, \& {Banchet}}]{Zagury}
{Zagury}, F., {Boulanger}, F., \& {Banchet}, V. 1999, \aap, 352, 645

\end{thebibliography}
\bibliographystyle{aasjournal}




\end{document}